\documentclass[aps,preprint]{revtex4}%
\usepackage{amsfonts}
\usepackage{amsmath}
\usepackage{amssymb}
\usepackage{graphicx}%
\providecommand{\U}[1]{\protect\rule{.1in}{.1in}}

\begin{document}
\title[ ]{The Classical Linear Oscillator in Classical Electrodynamics with Classical
Zero-Point Radiation}
\author{Timothy H. Boyer}
\affiliation{Department of Physics, City College of the City University of New York, New
York, New York 10031}

\begin{abstract}
In this article, we consider a classical charged particle in a one-dimensional
classical linear oscillator potential interacting with classical
electromagnetic zero-point radiation. \ We show that the oscillator has a
stable ground state and unstable resonant excited states which are analogous
to those of the quantum harmonic oscillator. \ We develop six themes. \ 1)
Classical electromagnetic zero-point radiation is Lorentz invariant and
requires that any mechanical system which comes to equilibrium in it is at
least approximately Lorentz invariant. \ Very few potentials meet this
approximately-Lorentz-invariant criterion. \ 2) Classical relativistic waves
travel at finite speed $c,$ which requires that the interaction of a point
charge $e$ located at $\mathbf{r}\left(  t\right)  $ be treated using the full
electric field $\mathbf{E}\left[  \mathbf{r}\left(  t\right)  ,t\right]  $,
not simply in the dipole approximation as $\mathbf{E}\left[  0,t\right]  $.
\ 3) The mechanical linear oscillator has $SO\left(  2\right)  $ symmetry
associated with the time behavior of the angle variable $\phi\left(  t\right)
=\omega_{0}t+\phi_{0}$, and leads to representations associated with integer
indices for the action variable $J$. \ 4) In the large-mass approximation, the
ground state is stable in classical zero-point radiation, even when higher
radiation multipoles are included in the analysis. \ 5) For the resonant
excited states, the predominant dipole radiation emitted by the charged
oscillator is balanced by the energy gained from higher frequencies of
classical electromagnetic zero-point radiation corresponding to odd multiples
of $\omega_{0}$. \ 6) The Bohr frequency condition $\Delta\mathcal{E=\hbar
\omega}_{0}$ appears naturally in a classical context. \ The presence of the
Lorentz-invariant zero-point radiation seems completely hidden in providing
the energy balance for the oscillating charged particle . \ The present
analysis puts a classical electromagnetic understanding under the old quantum
picture of electrons in classical orbits, but requires resonance of the
mechanical system in classical electromagnetic zero-point radiation. \ \ 

\end{abstract}
\maketitle

\section{Introduction}

\subsection{Summary}

In this article, we treat a nonrelativistic classical point charge $e$ in a
one-dimensional classical linear oscillator potential interacting with
classical electromagnetic zero-point radiation. \ We show that the oscillating
charge in the small-amplitude limit has a stable ground state and unstable
resonant excited states which are analogous to those of the quantum harmonic
oscillator. \ The present analysis provides a classical electromagnetic
understanding under the old-quantum picture of electrons in classical orbits.

\subsection{Basic Idea}

The purely nonrelativistic classical mechanical harmonic oscillator system
with parameters of mass $M$ and frequency $\omega_{0}$ can be written in terms
of action-angle variables, but there is no scale for the oscillator's action
variable \ $J_{e}$. \ On the other hand, random classical electromagnetic
zero-point radiation has exactly one scale, namely $\hbar$ associated with the
stochastic radiation action variable $J_{rad}$ for each radiation normal mode
where the average value is $\left\langle J_{rad}\right\rangle =\hbar/2$, and
the stochastic process is independent for each radiation normal mode..
\ Equilibrium between the charge and the zero-point radiation requires that
the mechanical oscillator receives the average value for its action variable
$J_{e}$ from the zero-point radiation $\hbar$. \ The action variable for the
$n^{th}$ resonant excited state $J_{e-n}$ of the mechanical system equals to
an odd multiple of the radiation action variable $J_{rad}$ as in
$J_{e-n}=\left(  2n+1\right)  J_{rad}$. \ Energy balance between the
mechanical system and the classical zero-point radiation\ is arranged by
differences in the frequencies of the driving radiation and the oscillator
motion, $\omega_{rad-n}=\left(  2n+1\right)  \omega_{0}$.

\subsection{Outline of the Article}

\ After reviewing some basic background material, we develop six fundamental
themes. \ 1) Classical electromagnetic zero-point radiation is Lorentz
invariant and requires that any charged mechanical system which comes to
equilibrium in it is at least approximately Lorentz invariant. \ Although the
linear harmonic oscillator is allowed as approximately Lorentz invariant, very
few other potentials $V\left(  \mathbf{r}\right)  $ are allowed.\ \ 2) We go
beyond the dipole approximation and require that the interaction of
relativistic electromagnetic waves $\mathbf{E}$ with a low-velocity charge at
$\mathbf{r}\left(  t\right)  $ be treated as $\mathbf{E}\left[  \mathbf{r}%
\left(  t\right)  ,t\right]  $, not simply in the dipole approximation
corresponding to $\mathbf{E}\left[  0,t\right]  $. \ 3) The charged mechanical
linear oscillator has $SO\left(  2\right)  $ symmetry associated with the
\textit{time} behavior of the angle variable, $\phi\left(  t\right)
=\omega_{0}t+\phi_{0}$, and the integer values for resonances arise from the
integer-indexed representations of the $SO\left(  2\right)  $ symmetry group.
\ 4) In the large-mass approximation, the ground state is stable even when
higher radiation multipoles are included. \ 5) For the resonant excited
states, the predominant dipole radiation emitted by the charged oscillator is
balanced by the energy picked up from higher frequencies of classical
electromagnetic zero-point radiation. \ 6) The Bohr frequency condition on
transitions between resonant excited states $\Delta\mathcal{E=\hbar\omega}%
_{0}$ appears naturally in the classical analysis. \ The presence of the
Lorentz-invariant zero-point radiation seems completely hidden in providing
the energy balance for the oscillating charged particle . \ 

\section{Background Material}

\subsection{Electromagnetic Model}

The classical one-dimensional charged harmonic oscillator can be considered to
arise in the classical electromagnetic model of the one-dimensional motion of
a charge $e$ between two fixed charges of the same sign as $e$. \ Then the
moving charge $e$ oscillates back and forth, being repelled by both of the
fixed charges. \ Since the oscillating charge is oscillating and hence is
accelerating, it loses energy through the emission of radiation. \ The charge
$e$ would eventually come to rest after having lost all its mechanical energy
if it were not for the ambient classical zero-point radiation which provides
random forces to accelerate the charge $e$. \ We want to consider the behavior
of the oscillating charge in the classical zero-point radiation spectrum. \ We
will choose the charged oscillator as oriented along the $z$-axis so that the
spherical angle is $\theta=0$, and there is no need to discuss the azimuthal angle.

\subsection{Lorentz Invariance, at Least in Approximation}

The fundamental constants of electromagnetic theory are the elementary charge
$e$, the speed of light $c$, and Planck's constant $\hbar$. \ Thus there is
room for one \textit{unit} mass, $M_{u},$ which will set the scale for the
theory. \ The unit of length can be taken as $e^{2}/\left(  M_{u}c^{2}\right)
$, the unit of time as $e^{2}/\left(  M_{u}c^{3}\right)  $, and the unit of
energy as $M_{u}c^{2}$. \ \ The fine structure constant $e^{2}/\left(  \hbar
c\right)  $ is the fundamental constant independent of the choice of unit
mass. \ In the present article, the speed of light $c$ does not enter the
\textit{mechanical} motion of the low-velocity charge $e,$ and hence the
\textit{only} appearance of $c$ is in the connection of the charged mechanical
system to radiation. \ Therefore the fine structure constant does not appear
in the analysis. \ Also, the charge $e$ appears for both the energy lost and
the energy gained by the oscillator, and so cancels out completely in the
equations for oscillator energy balance. \ In the nonrelativistic analysis,
the constant $c$ does not appear in the mechanical motion, and the charge $e$
can assume any (small) value. \ 

\subsection{Classical Electrodynamics with Classical Zero-Point Radiation}

The existence of Casimir forces\cite{Casimir} between uncharged conducting
surfaces implies, within classical physics, that there must be random
radiation even at the zero of temperature, termed classical electromagnetic
zero-point radiation.\cite{BCas}\cite{CasimirForces}\ The magnitude and
distance behavior of the Casimir forces can be accurately accounted for by
temperature-independent random classical radiation with a spectrum per normal
mode given by an \textit{average} energy $U_{rad}\left(  \omega\right)
=\hbar\omega/2,$ which is the spectrum of Lorentz-invariant classical
zero-point radiation. \ We emphasize that the instantaneous radiation energy
per normal mode is $J_{rad}\omega$ where the action variable $J_{rad}$ is a
stochastic process. \ The action variable for classical zero-point radiation
has the same average value $\left\langle J_{rad}\right\rangle =\hbar/2$ for
radiation of every frequency $\omega$ in every inertial frame.\ 

Classical electrodynamics with classical electromagnetic zero-point radiation
is a specific version of classical electrodynamics where the source-free
solution of Maxwell's equations is chosen as random classical radiation fields
with a Lorentz-invariant energy spectrum.\cite{Marshall}\cite{B1975} \ The
theory is often termed \textquotedblleft stochastic
electrodynamics.\textquotedblright\ \ The scale of the classical zero-point
radiation is chosen so as to give correctly the Casimir forces between
conducting parallel plates. \ This \textit{classical} theory contains Planck's
constant $\hbar$ as the \textit{one and only} \textit{scale} for classical
zero-point radiation. \ For over 50 years, the implications of this classical
theory have been gradually obtained. \ The \textit{classical} theory has given
a number of results which are usually claimed to require quantum theory. \ For
example, there are classical calculations for Casimir forces, van der Waals
force, oscillator specific heats, diamagnetism, superfluid behavior, the
absence of atomic collapse,\cite{ColeZou} and the blackbody spectrum\cite{BBc}
which agree for average values with the results of experiment and with the
corresponding quantum calculations.\cite{B2019}

\subsection{Approximately Relativistic Behavior for Large Oscillator Mass}

Classical electromagnetic zero-point radiation is Lorentz
invariant.\cite{Marshall} \ Any \textit{nonrelativistic} system, such as a
\textit{nonlinear} charged harmonic oscillator with arbitrary oscillation
speed, will tend to push the Lorentz-invariant zero-point spectrum toward the
Rayleigh-Jeans spectrum.\cite{Bnon} \ However, a nonlinear oscillator is
\textit{not} even approximately Lorentz invariant because it contains a length
parameter in the nonlinear term which can be combined with the natural
oscillator frequency $\omega_{0}$ to give a parameter-based velocity $v$ which
is different from the speed of light $c$. \ For approximate Lorentz symmetry,
we turn here to a \textit{linear} oscillator which involves parameters of a
mass $M$ and a frequency $\omega_{0}$. \ Furthermore, it is only when the
charged oscillator amplitude $Z_{0}\left(  J_{e}\right)  $ (depending on the
oscillator's action variable $J_{e}$) is very small, and therefore its maximum
speed $v=Z_{0}\omega_{0}$ is very small, $v<<c,$ that the linear oscillator
motion is approximately in agreement with a relativistic system. \ Thus, we
will insist that the motion of the charged particle $e$ is small by taking its
mass $M$ as very large so that the maximum oscillator kinetic energy is small,
$M\left(  Z_{0}\omega_{0}\right)  ^{2}/2=Mv^{2}/2<<Mc^{2}$, giving $v<<c$.
\ The emphasis on \textit{relativity} sharply restricts the mechanical systems
to which the theory applies. \ The one-dimensional linear harmonic oscillator
and the Coulomb potential are the most important allowed mechanical potential systems.

\subsection{Oscillator Ground State in the Present Analysis}

It is important to realize that, when treated in the \textit{dipole}
approximation, the charged linear oscillator comes to steady state with
\textit{any} spectrum of random classical radiation,\cite{B1975} as does a
nonrelativistic charged particle in a Coulomb potential. \ However, the
assumption, that the oscillator (in the limiting small-amplitude oscillation)
comes to equilibrium for \textit{both }its dipole and also its quadrupole
radiation fields, fixes the equilibrium spectrum (up to an overall constant)
as Lorentz-invariant classical zero-point radiation.\cite{BComm} \ In its
\textit{ground} state, the interaction of the charged one-dimensional linear
harmonic oscillator with radiation is completely disguised. \ The oscillator
scatters the zero-point radiation, but there is no time-average net radiation
propagating in any direction.\cite{B1975} \ One might be tempted to conclude
that the charged harmonic oscillator had no interaction with radiation despite
its continued oscillation. \ 

\subsection{Resonant Excited States in the Present Analysis}

\textit{All} frequencies are present in random classical zero-point radiation.
\ On the other hand, the oscillating \textit{mechanical} system has only
\textit{one} natural oscillating frequency $\omega_{0}$. \ If the charged
mechanical oscillator is in its ground state in the presence of classical
zero-point radiation, the radiation at both the dipole and quadrupole
frequencies gives rise to the \textit{same }stochastic process for the
mechanical oscillator. \ An observer measuring radiation would be unaware of
the radiation continually emitted and absorbed by the charged mechanical
oscillator. \ However, if the mechanical oscillator is well above the
amplitude of the ground state, the charged oscillator is still emitting
radiation, predominantly into the dipole radiation mode at the same frequency
as its natural mechanical oscillation frequency $\omega_{0}$. \ If the
oscillator amplitude is correct for \textit{resonance} with some driving
zero-point radiation mode, the dipole radiation emission will be balanced
against the dipole energy gained from zero-point radiation at a frequency
\textit{different} from the natural oscillation frequency $\omega_{0}$ of the
mechanical oscillator. \ For these resonant excited states, the
\textit{dipole} radiation emitted by the charged oscillator at its fundamental
frequency $\omega_{0}$ with stochastic action variable $J_{e-n}=\left(
2n+1\right)  J_{rad}$ can be balanced by the \textit{dipole} energy gain from
zero-point radiation modes at a frequency $\omega_{rad-n}=\left(  2n+1\right)
\omega_{0}$ different from the natural mechanical oscillation frequency
$\omega_{0}$ of the oscillator. \ However, the \textit{energy }$\mathcal{E}%
_{e-n}$\textit{ }of the oscillator $J_{e-n}\omega_{0}=\left[  \left(
2n+1\right)  J_{e-1}\right]  \omega_{0}=J_{rad}\left[  \left(  2n+1\right)
\omega_{0}\right]  =J_{rad}\omega_{rad-n}$ is the same as the \textit{energy}
$\mathcal{E}_{rad-n}$ of the associated zero-point radiation. \ On changes of
resonant states $\Delta\mathcal{E}_{e}$, the situation leads to Bohr's
condition $\Delta\mathcal{E}_{e}\mathcal{=\hbar}\omega_{0}$. \ In the
remainder of this article, we will carry out the calculations to confirm these statements.

\section{The Linear Oscillator System}

The one-dimensional classical linear harmonic oscillator oriented along the
$z$-axis has a Hamiltonian given by
\begin{equation}
H\left(  p,z\right)  =\frac{p^{2}}{2M}+\frac{1}{2}M\omega_{0}^{2}z^{2}
\label{SHO}%
\end{equation}
where the position $z$ and linear momentum $p$ are the dynamical variables on
phase space, and the mass $M$ and oscillation (angular) frequency $\omega_{0}$
are fixed parameters. \ The Hamiltonian $H$ may be rewritten in terms of
action-angle variables $J,$ and $\phi$ as\cite{Goldstein462}
\begin{equation}
H\left(  J,\phi\right)  =J\omega_{0},
\end{equation}
where the angle variable $\phi=\omega_{0}t+\phi_{0}$ does not appear, nor does
the mass $M$. \ The solution $z\left(  t\right)  $ of the equation of motion
from this Hamiltonian, written as a \textit{multiply periodic} expansion,
involves absolutely no \textit{harmonics} of the \textit{mechanical}
oscillation frequency $\omega_{0}$, so that\cite{Goldstein462}%

\begin{equation}
z\left(  t\right)  =Z_{0}\sin\left[  \phi\left(  t\right)  \right]
=\sqrt{\frac{2J}{M\omega_{0}}}\sin\left[  \phi\left(  t\right)  \right]
=\sqrt{\frac{2J}{M\omega_{0}}}\sin\left[  \omega_{0}t+\phi_{0}\right]
\label{ztZ0}%
\end{equation}
and
\begin{equation}
p\left(  t\right)  =M\omega_{0}Z_{0}\cos\left[  \phi\left(  t\right)  \right]
=\sqrt{2M\omega_{0}J}\cos\left[  \phi\left(  t\right)  \right]  =\sqrt
{2M\omega_{0}J}\cos\left[  \omega_{0}t+\phi_{0}\right]  ,
\end{equation}
where the amplitude of the motion is given by
\begin{equation}
Z_{0}=\sqrt{\frac{2J}{M\omega_{0}}}. \label{Z0}%
\end{equation}
Thus, in time, the mechanical system undergoes a uniform rotation in the angle
$\phi=\omega_{0}t+\phi_{0}$ on the two-dimensional phase space labelled by $p$
and $z$ or by $J$ and $\phi$. \ These orbits correspond to the ellipses often
mentioned in texts of statistical mechanics.\cite{Reif55} \ The Hamiltonian
may be said to have an $SO\left(  2\right)  $ symmetry on the two-dimensional
phase space.\cite{Miller}\ \ 

The action variable $J$ for the oscillator is indeed given by%
\begin{align}
\frac{1}{2\pi}\oint pdz  &  =\frac{1}{2\pi}\oint\sqrt{2M\omega_{0}J}%
\cos\left[  \omega_{0}t+\phi_{0}\right]  d\left\{  \sqrt{\frac{2J}{M\omega
_{0}}}\sin\left[  \omega_{0}t+\phi_{0}\right]  \right\} \nonumber\\
&  =\frac{1}{2\pi}\int_{0}^{2\pi/\omega_{0}}2J\omega_{0}\cos^{2}\left[
\omega_{0}t+\phi_{0}\right]  dt=J.
\end{align}
The oscillator energy $\mathcal{E}$ is given by%
\begin{equation}
\mathcal{E=}\frac{p^{2}}{2M}+\frac{1}{2}M\omega_{0}^{2}z^{2}=\frac{1}%
{2}M\omega_{0}^{2}Z_{0}^{2}=J\omega_{0}.
\end{equation}

\section{The Spherical Mode Expansion for Random Classical Radiation}

The discrete integer values for certain parameters are associated with the
irreducible representations of groups of symmetries. \ In previous analyses of
classical electrodynamics with classical electromagnetic zero-point radiation,
plane waves were used, which fitted with the emphasis on Lorentz
invariance.\cite{B1975} \ For the random classical electromagnetic radiation
in the present article, we will emphasize the \textit{rotational} symmetry
involving the rotation groups $SO\left(  2\right)  $ and/or $SO\left(
3\right)  $.\cite{Zee} \ Since the electromagnetic waves are in three spatial
dimensions, we expand in terms of spherical multipole radiation fields. \ Then
random radiation in a very large \textit{spherical} cavity of radius
$\mathsf{R}$ can be written as\cite{Jackson}%

\begin{align}
\mathbf{E}(\mathbf{r,}t)  &  =\operatorname{Re}\sum\nolimits_{n=1}^{\infty
}\sum\nolimits_{l=1}^{\infty}\sum\nolimits_{m=-l}^{m=l}\left\{  \exp\left[
i\left(  -k_{nl}^{M}ct+\theta_{nlm}^{M}\right)  \right]  \left[  ia_{nlm}%
^{M}j_{l}\left(  k_{nlm}^{M}r\right)  \mathbf{X}_{l,m}\left(  \theta
,\phi\right)  \right]  \right. \nonumber\\
&  \left.  +\exp\left[  i\left(  -k_{nlm}^{E}ct+\theta_{nlm}^{E}\right)
\right]  \left[  a_{nlm}^{E}/\left(  -ik_{nlm}^{E}\right)  \right]
\nabla\times\left[  j_{l}\left(  k_{nlm}^{E}r\right)  \mathbf{X}_{lm}\left(
\theta,\phi\right)  \right]  \right\}  , \label{Ezprt}%
\end{align}
and%

\begin{align}
\mathbf{B}(\mathbf{r,}t)  &  =\operatorname{Re}\sum\nolimits_{n=1}^{\infty
}\sum\nolimits_{l=1}^{\infty}\sum\nolimits_{m=-l}^{m=l}\left\{  \exp\left[
i\left(  -k_{nlm}^{E}ct+\theta_{nlm}^{E}\right)  \right]  \left[  ia_{nlm}%
^{E}j_{l}\left(  k_{nlm}^{E}r\right)  \mathbf{X}_{l,m}\left(  \theta
,\phi\right)  \right]  \right. \nonumber\\
&  \left.  +\exp\left[  i\left(  -k_{nlm}^{M}ct+\theta_{nlm}^{M}\right)
\right]  \left[  a_{nlm}^{M}/\left(  ik_{nlm}^{M}\right)  \right]
\nabla\times\left[  j_{l}\left(  k_{nlm}^{M}r\right)  \mathbf{X}_{lm}\left(
\theta,\phi\right)  \right]  \right\}  , \label{Bzprt}%
\end{align}
where the parameters $a_{lm}^{E}\left(  k_{nl}^{E}\right)  $ and $a_{lm}%
^{M}\left(  k_{nl}^{M}\right)  $ are associated with the particular spectrum
of random electromagnetic radiation, $\mathbf{X}_{lm}\left(  \theta
,\phi\right)  $ is a vector spherical harmonic, and the random phases
$\theta_{nlm}^{E}$ and $\theta_{nlm}^{M}$ are distributed uniformly on
$[0,2\pi)$ and independently for each radiation mode.\cite{EHR} \ For a
general spectrum of random radiation, we have the scale given by\cite{B2022}%

\begin{equation}
\left\vert a_{nlm}^{E}\right\vert ^{2}=\frac{16\pi\left(  k_{nl}^{E}\right)
^{2}}{\mathsf{R}}\left[  U_{nlm}^{E}\left(  k_{nlm}^{E}\right)  \right]
=\frac{16\pi\left(  \omega_{nl}^{E}\right)  ^{3}}{c^{2}\mathsf{R}}\left\langle
\left[  J_{rad-nlm}\left(  \omega\right)  \right]  \right\rangle , \label{aE}%
\end{equation}
with the stochastic average electric energy $U_{nlm}^{E}$ of the radiation
normal mode given by $U_{nlm}^{E}\left(  k_{nlm}^{E}\right)  =k_{nlm}%
^{E}c\left\langle \left[  J_{rad~nlm}\left(  \omega\right)  \right]
\right\rangle _{\theta_{nlm}}=\omega_{nlm}^{E}\left\langle \left[
J_{rad~nlm}\left(  \omega\right)  \right]  \right\rangle _{\theta_{nlm}}$,
where $\left\langle \left[  J_{rad~nlm}\left(  \omega\right)  \right]
\right\rangle _{\theta_{nlm}}$ means the stochastic average of the action
variable $J_{rad-nlm}$ which is a function of $\omega.$ \ \ There is an
expression analogous to Eq. (\ref{aE}) for the magnetic radiation modes
$\left\vert a_{nlm}^{M}\right\vert ^{2}$. \ \ However, there is no
\textit{magnetic}-mode radiation coupling to the straight-line motion of a
one-dimensional charged linear harmonic oscillator along the $z$-axis, so that
only the \textit{electric} modes drive the oscillator. \ \ Here we use random
phases for the stochastic character.\ The number of normal modes per unit
(angular) frequency per unit volume is $\omega^{2}/\left(  \pi^{2}%
c^{3}\right)  $ which gives a $\theta_{nlm}^{E}$- and $\theta_{nlm}^{M}%
$-average radiation energy per unit angular frequency interval per unit volume
$\left[  \omega^{2}/\left(  \pi^{2}c^{3}\right)  \right]  U\left(
\omega\right)  .$ \ 

For the limit of large radius $\mathsf{R}$ of the enclosing sphere containing
standing waves, we have\cite{B2022} $dk=\pi dn/\mathsf{R}$ and%
\begin{equation}
\sum\nolimits_{n=1}^{\infty}\rightarrow\int_{0}^{\infty}dn=\int_{0}^{\infty
}dk\frac{\mathsf{R}}{\pi}=\int_{0}^{\infty}d\omega\frac{\mathsf{R}}{\pi c}.\,
\label{SumC}%
\end{equation}

\section{Energy Absorbed From Random Ambient Radiation}

\subsection{Energy Balance for Both Dipole and Quadrupole Radiation}

If the classical charged oscillator were located in absolutely empty infinite
space, the emission of classical radiation would occur until all the
oscillator's mechanical energy was transferred to electromagnetic radiation.
\ However, if the oscillator is located in a field of random classical
radiation, such as thermal radiation, the oscillator will also gain energy
from the ambient radiation. \ A steady-state situation is reached when, on
average, the energy emitted by the charged oscillator is balanced by the
average energy gained from the random ambient radiation. \ Such a balance was
first calculated (\textit{in the dipole approximation only}) by Planck at the
end of the nineteenth century.\cite{Planck} \ Planck concluded that the linear
oscillator came to equilibrium when the oscillator time-average energy matched
the average energy of the modes in the radiation field at the frequency
corresponding to the natural frequency of the mechanical oscillator.

For the oscillator ground state, we will go beyond the \textit{dipole}
approximation and will consider also \textit{quadrupole} radiation in the
spherical multipole radiation field with which the oscillator interacts in the
small-source approximation. \ In an earlier article,\cite{BComm} we showed
that if the radiation spectrum was stable for both dipole and quadrupole
scattering by a small one-dimensional harmonic oscillator and no
velocity-dependent damping was assumed, then the \textit{only} allowed
spectrum was that of classical electromagnetic zero-point radiation where the
action variable is independent of the frequency. \ In the earlier article, the
classical zero-point radiation was expanded using plane waves; here we will
use the spherical mode expansion in Eqs. (\ref{Ezprt})\ and (\ref{Bzprt}).
\ We will assume that classical zero-point radiation with average energy
$U_{rad}\left(  \omega\right)  =\hbar\omega/2$ is involved for the radiation
spectrum, and we will calculate the steady-state situation using the
variation-of-parameters technique.\cite{Born424}\cite{Greenberg141}

\subsection{Beyond the Usual Approximation Involving Dipole Radiation Only}

The usual \textit{dipole} approximation for the charged linear oscillator
takes forcing by an electric field as $eE_{z}\left(  0,t\right)  $, thus$\,$
neglecting any displacement of the charge and writing\cite{Born424}%

\begin{equation}
M\,\ddot{z}=-M\omega_{0}^{2}z\left(  t\right)  +eE_{z}(0\mathbf{,}t),
\end{equation}
where the loss of energy due to radiation emission is considered separately.
\ In this equation, the linear harmonic oscillator has exactly one resonant
frequency at $\omega_{0}$. \ If the oscillator is initially at rest, the
driving field $E_{z}\left(  0,t\right)  $ will cause the particle to gain
energy. \ If the driving field is random radiation, the linear oscillator will
(on average) gain enough energy to balance the loss of mechanical energy due
to emission into dipole radiation at an energy $\mathcal{E}=\left(
1/2\right)  \hbar\omega_{0}$. \ This is the basis of Planck's result, that the
average energy of the oscillator will equal the average energy per normal mode
of the radiation field at the frequency equal to the natural frequency
$\omega_{0}$ of the linear oscillator. \ This approximate analysis does not
compare the speed $\dot{z}$ of the oscillator with the speed $c$ of the random
radiation, nor does it allow for any resonant interaction with higher
frequency radiation. \ The present article goes beyond this dipole
approximation and uses $\mathbf{E}\left[  \mathbf{r}\left(  t\right)
,t\right]  $ as the driving force for the charged linear oscillator. \ Note
the position-dependence $\mathbf{r}\left(  t\right)  $ included in the driving radiation.

\subsection{Variation of Parameters}

For the gain of power by the charged mechanical oscillator from the random
zero-point radiation, we require that there exists a time $\tau$ long enough
so that there are many cycles of oscillation during $\tau,$ but short enough
such that the amplitude $Z_{0}$ of the oscillator does not change
significantly during $\tau$.\cite{MarshallP} \ For a small charge $e$ on the
oscillator, such a situation is indeed possible, since small charge $e$ means
that little energy is gained or lost during each oscillation cycle. \ For the
large-mass $M$ (nonrelativistic) approximation (but still ignoring the
emission of radiation energy), we have Newton's second law for a
nonrelativistic particle of mass $M$%
\begin{equation}
M\,\ddot{z}=-M\omega_{0}^{2}z\left(  t\right)  +eE_{z}\left[  \widehat{z}%
z\left(  t\right)  \mathbf{,}t\right]  , \label{Mzdd}%
\end{equation}
where $\mathbf{r}\left(  t\right)  =\widehat{z}z\left(  t\right)  $ gives the
mechanical motion of the oscillator. \ Note particularly the dependence of the
force on the position $\widehat{z}z\left(  t\right)  $ in the driving
radiation. \ This aspect will be important for any higher multipoles,
including the resonant excited states. \ 

\subsection{Solution of the Differential Equation}

We are interested in the particular solution $z_{p}\left(  t\right)  $ given
by the variation of parameters\cite{Greenberg141},
\begin{equation}
z_{p}\left(  t\right)  =\frac{e}{M\omega_{0}}\int_{0}^{t}dt^{\prime}%
E_{z}\left[  \mathbf{r}\left(  t^{\prime}\right)  \mathbf{,}t^{\prime}\right]
\sin\left[  \omega_{0}\left(  t-t^{\prime}\right)  \right]  , \label{zpt}%
\end{equation}
where $\mathbf{r}\left(  t\right)  =\widehat{z}z\left(  t\right)  $. \ This
expression (\ref{zpt}) satisfies Eq. (\ref{Mzdd}) when we calculate the
velocity and acceleration of the particle. \ We have the velocity of the
charge $e$
\begin{equation}
\frac{dz_{p}}{dt}=\frac{e}{M}\int_{0}^{t}dt^{\prime}E_{z}\left[
\mathbf{r}\left(  t^{\prime}\right)  \mathbf{,}t^{\prime}\right]  \cos\left[
\omega_{0}\left(  t-t^{\prime}\right)  \right]  , \label{ddtzp}%
\end{equation}
and acceleration%
\begin{equation}
\frac{d^{2}z_{p}\left(  t\right)  }{dt^{2}}=-\frac{e\omega_{0}}{M}\int_{0}%
^{t}dt^{\prime}E_{z}\left[  \mathbf{r}\left(  t^{\prime}\right)
\mathbf{,}t^{\prime}\right]  \sin\left[  \omega_{0}\left(  t-t^{\prime
}\right)  \right]  +\frac{e}{M}E_{z}\left[  \mathbf{r}\left(  t^{\prime
}\right)  \mathbf{,}t^{\prime}\right]  .
\end{equation}
Thus, the expression for $z_{p}\left(  t\right)  $ in Eq. (\ref{zpt})
satisfies the differential equation (\ref{Mzdd}) and also the initial
conditions $z_{p}\left(  0\right)  =0$ and $\dot{z}_{p}\left(  0\right)  =0$. \ 

\subsection{Energy Gain During Time $\tau$}

The energy gained from the random radiation by the oscillator during the time
$\tau$ is given by the time integral of the power delivered. \ If we include
the position-dependence of the electromagnetic driving field, we have from Eq.
(\ref{ddtzp})
\begin{align}
W\left(  \tau\right)   &  =\int_{0}^{\tau}dt\left[  \frac{d}{dt}z_{p}\left(
t\right)  \right]  eE_{z}\left[  \mathbf{r}\left(  t\right)  \mathbf{,}%
t\right] \nonumber\\
&  =\int_{0}^{\tau}dt\left[  \frac{e}{M}\int_{0}^{t^{\prime}=t}dt^{\prime
}E_{z}(\mathbf{r}\left(  t^{\prime}\right)  \mathbf{,}t^{\prime})\cos\left[
\omega_{0}\left(  t-t^{\prime}\right)  \right]  \right]  eE_{z}\left[
\mathbf{r}\left(  t\right)  \mathbf{,}t\right] \nonumber\\
&  =\frac{e^{2}}{M}\int_{0}^{\tau}dt\int_{0}^{t^{\prime}=t}dt^{\prime}%
\cos\left[  \omega_{0}\left(  t-t^{\prime}\right)  \right]  E_{z}\left[
\mathbf{r}\left(  t^{\prime}\right)  \mathbf{,}t^{\prime}\right]  E_{z}\left[
\mathbf{r}\left(  t\right)  \mathbf{,}t\right]  . \label{Wtau}%
\end{align}
Now the integrand of the double integral in Eq. (\ref{Wtau}) is symmetric
under interchange of $t$ and $t^{\prime}$. \ The region of integration is an
isosceles right triangle having axes labeled by $t$ and $t^{\prime}$, with the
integration in $t^{\prime}$ first running for $0$ to $t$, and then the
integral in $t$ running from $0$ to $\tau$. \ However, since the integrand is
symmetric, one can reverse the order of integration and integrate first in $t$
from $t^{\prime}$ to $\tau$, and then in $t^{\prime}$ from $0$ to $\tau$
giving\cite{Born424}\cite{Reif571}%
\begin{equation}
W\left(  \tau\right)  =\int_{0}^{\tau}dt^{\prime}\int_{t^{\prime}}^{\tau
}dt^{\prime}\frac{e^{2}}{M}\cos\left[  \omega_{0}\left(  t-t^{\prime}\right)
\right]  E_{z}\left[  \mathbf{r}\left(  t^{\prime}\right)  \mathbf{,}%
t^{\prime}\right]  E_{z}\left[  \mathbf{r}\left(  t\right)  \mathbf{,}%
t\right]  .
\end{equation}
But then we can interchange the labels on $t$ and $t^{\prime}$ and add half of
each of the two expressions to obtain%
\begin{equation}
W\left(  \tau\right)  =\frac{1}{2}\int_{0}^{\tau}dt\int_{0}^{\tau}dt^{\prime
}\frac{e^{2}}{M}\cos\left[  \omega_{0}\left(  t-t^{\prime}\right)  \right]
E_{z}\left[  \mathbf{r}\left(  t^{\prime}\right)  \mathbf{,}t^{\prime}\right]
E_{z}\left[  \mathbf{r}\left(  t\right)  \mathbf{,}t\right]  . \label{Wtauh}%
\end{equation}

\subsection{Energy Absorbed from Random Radiation}

Random radiation involves a stochastic process in time which may be
described\cite{EHR} using random phases $\theta_{nlm}^{E}$. \ The radiation
interacts with the charged dipole oscillator along the $z$-axis. \ What is of
interest to us is the stochastic average values involving averages over
$\theta_{n1,0}^{E}$. \ The averages over the random phases of the zero-point
radiation, give%
\begin{equation}
\left\langle \cos\left[  \theta_{n1,0}^{E}\right]  \cos\left[  \theta
_{n^{\prime}1,0}^{E}\right]  \right\rangle _{\theta^{E}}=\left\langle
\sin\left[  \theta_{n1,0}^{E}\right]  \sin\left[  \theta_{n^{\prime}1,0}%
^{E}\right]  \right\rangle _{\theta^{E}}=\frac{1}{2}\delta_{nn^{\prime}}%
\end{equation}
and%
\begin{equation}
\left\langle \cos\left[  \theta_{n1,0}^{E}\right]  \sin\left[  \theta
_{n^{\prime}1,0}^{E}\right]  \right\rangle _{\theta^{E}}=0.
\end{equation}
Thus, averaging over the random phases and then summing over the Kronecker
delta in $(1/2)\delta_{n,n^{\prime}}$, we obtain
\begin{align}
&  \left\langle E_{z}\left[  \mathbf{r}\left(  t^{\prime}\right)
\mathbf{,}t^{\prime}\right]  E_{z}\left[  \mathbf{r}\left(  t\right)
\mathbf{,}t\right]  \right\rangle _{\theta^{E}}\nonumber\\
&  =\left\langle \sum\nolimits_{nl}E_{z}^{nl0}\left[  \mathbf{r}\left(
t\right)  ,0\right]  \cos\left[  -\omega_{nl}^{E}t+\theta_{nlm}^{E}\right]
\sum\nolimits_{n^{\prime}l^{\prime}}E_{z}^{n^{\prime}l^{\prime}0}\left[
\mathbf{r}\left(  t^{\prime}\right)  ,0\right]  \cos\left[  -\omega_{nl}%
^{E}t^{\prime}+\theta_{n^{\prime}l^{\prime}m^{\prime}}^{E}\right]
\right\rangle _{\theta^{E}}\nonumber\\
&  =\sum\nolimits_{nl}\frac{1}{2}\cos\left[  -\omega_{nlm}^{E}\left(
t-t^{\prime}\right)  \right]  \left\{  E_{z}^{nl0}\left[  \mathbf{r}\left(
t\right)  ,0\right]  E_{z}^{nl0}\left[  \mathbf{r}\left(  t^{\prime}\right)
,0\right]  \right\}  . \label{EzEztt}%
\end{align}

Then from Eq. (\ref{EzEztt}), we have from Eq. (\ref{Wtauh}) for the average
energy gained from the random radiation
\begin{align}
&  \left\langle W\left(  \tau\right)  \right\rangle _{\theta^{E}}\nonumber\\
&  =\frac{e^{2}}{2M}\int_{0}^{\tau}dt\int_{0}^{\tau}dt^{\prime}\cos\left[
\omega_{0}\left(  t-t^{\prime}\right)  \right]  \left\langle E_{z}\left[
\mathbf{r}\left(  t\right)  \mathbf{,}t\right]  E_{z}\left[  \mathbf{r}\left(
t^{\prime}\right)  \mathbf{,}t^{\prime}\right]  \right\rangle _{\theta^{E}%
}\nonumber\\
&  =\frac{e^{2}}{2M}\int_{0}^{\tau}dt\int_{0}^{\tau}dt^{\prime}\cos\left[
\omega_{0}\left(  t-t^{\prime}\right)  \right]  \sum\nolimits_{n=1}^{\infty
}\frac{1}{2}\cos\left[  -k_{nl}^{E}c\left(  t-t^{\prime}\right)  \right]
E_{z}^{nl0}\left[  \mathbf{r}\left(  t\right)  ,0\right]  E_{z}^{nl0}\left[
\mathbf{r}\left(  t^{\prime}\right)  ,0\right] \nonumber\\
&  =\frac{e^{2}}{2M}\sum\nolimits_{n=1}^{\infty}\int_{0}^{\tau}dt\int%
_{0}^{\tau}dt^{\prime}\frac{1}{2}\left\{  E_{z}^{nl}\left[  \mathbf{r}\left(
t\right)  ,0\right]  E_{z}^{nl}\left[  \mathbf{r}\left(  t^{\prime}\right)
,0\right]  \right\} \nonumber\\
&  \times\frac{1}{2}\left\{  \cos\left[  \left(  \omega_{nlm}^{E}-\omega
_{0}\right)  \left(  t-t^{\prime}\right)  \right]  +\cos\left[  \left(
\omega_{nlm}^{E}+\omega_{0}\right)  \left(  t-t^{\prime}\right)  \right]
\right\}
\end{align}
Also, we use the expansions
\begin{align}
&  \cos\left[  \left(  \omega_{nlm}^{E}\pm\omega_{0}\right)  \left(
t-t^{\prime}\right)  \right] \nonumber\\
&  =\cos\left[  \left(  \omega_{nlm}^{E}\pm\omega_{0}\right)  t\right]
\cos\left[  \left(  \omega_{nlm}^{E}\pm\omega_{0}\right)  t^{\prime}\right]
\mp\sin\left[  \left(  \omega_{nlm}^{E}\pm\omega_{0}\right)  t\right]
\sin\left[  \left(  \omega_{nlm}^{E}\pm\omega_{0}\right)  t^{\prime}\right]
\end{align}
and reorganize the terms to obtain%

\begin{align}
\left\langle W\left(  \tau\right)  \right\rangle _{\theta^{E}}  &
=\frac{e^{2}}{8M}\sum\nolimits_{n=1}^{\infty}\int_{0}^{\tau}dt\int_{0}^{\tau
}dt^{\prime}\left\{  E_{z}\left[  \mathbf{r}\left(  t\right)  ,0\right]
E_{z}\left[  \mathbf{r}\left(  t^{\prime}\right)  ,0\right]  \right\}
\cos\left[  \left(  \omega_{nlm}^{E}-\omega_{0}\right)  \left(  t-t^{\prime
}\right)  \right] \nonumber\\
&  =\frac{e^{2}}{8M}\sum\nolimits_{n=1}^{\infty}\left[  \int_{0}^{\tau}%
dtE_{z}\left[  \mathbf{r}\left(  t\right)  ,0\right]  \cos\left[  \left(
\omega_{nlm}^{E}-\omega_{0}\right)  t\right]  \right]  ^{2}\nonumber\\
&  +\frac{e^{2}}{8M}\sum\nolimits_{n=1}^{\infty}\left[  \int_{0}^{\tau}%
dtE_{z}\left[  \mathbf{r}\left(  t\right)  ,0\right]  \sin\left[  \left(
\omega_{nlm}^{E}-\omega_{0}\right)  t\right]  \right]  ^{2}\nonumber\\
&  +\frac{e^{2}}{8M}\sum\nolimits_{n=1}^{\infty}\left[  \int_{0}^{\tau}%
dtE_{z}\left[  \mathbf{r}\left(  t\right)  ,0\right]  \cos\left[  \left(
\omega_{nlm}^{E}+\omega_{0}\right)  t\right]  \right]  ^{2}\nonumber\\
&  +\frac{e^{2}}{8M}\sum\nolimits_{n=1}^{\infty}\left[  \int_{0}^{\tau}%
dtE_{z}\left[  \mathbf{r}\left(  t\right)  ,0\right]  \sin\left[  \left(
\omega_{nlm}^{E}+\omega_{0}\right)  t\right]  \right]  ^{2}. \label{Wtau8}%
\end{align}

\subsection{Energy Gained from Random Classical Radiation}

The full driving term is obtained from Eq. (\ref{Ezprt}) as the
\textit{radial} component of $\mathbf{E}(\mathbf{r,}t)$ at $\theta=0$ and
$m=0$ giving\cite{Jackson745}%
\begin{equation}
E_{z}(\mathbf{r}\left(  t\right)  \mathbf{,}0)=-\sum\nolimits_{nl}\frac
{a_{l0}^{E}}{\sqrt{l\left(  l+1\right)  }}l\left(  l+1\right)  \frac
{j_{l}\left[  k\left\vert z\left(  t\right)  \right\vert \right]
}{k\left\vert z\left(  t\right)  \right\vert }Y_{l,0}\left(  \theta
=0,0\right)  \label{Ezdlm}%
\end{equation}
where we have separated out the part involving the time and random phases.
\ Then we have from Eqs. (\ref{Ezdlm}), (\ref{aE}), and (\ref{Wtau8})
\begin{align}
&  \left\langle W\left(  \tau\right)  \right\rangle _{\theta^{E}}=\frac{e^{2}%
}{8M}\sum\nolimits_{nl}\left[  \frac{16\pi\left(  \omega_{nl}^{Ed}\right)
^{3}}{c^{2}\mathsf{R}}\left\langle J_{rad}\right\rangle \right]  l\left(
l+1\right)  \left\vert Y_{l,0}\right\vert ^{2}\nonumber\\
&  \times\left\{  \left[  \int_{0}^{\tau}dt\frac{j_{l}\left(  k_{nl}%
^{E}\left\vert z\left(  t\right)  \right\vert \right)  }{k_{nl}^{E}\left\vert
z\left(  t\right)  \right\vert }\cos\left[  \left(  \omega_{nl}^{E}-\omega
_{0}\right)  t\right]  \right]  ^{2}+\left[  \int_{0}^{\tau}dt\frac
{j_{l}\left(  k_{nl}^{E}\left\vert z\left(  t\right)  \right\vert \right)
}{k_{nl}^{E}\left\vert z\left(  t\right)  \right\vert }\sin\left[  \left(
\omega_{nl}^{E}-\omega_{0}\right)  t\right]  \right]  ^{2}\right. \nonumber\\
&  +\left.  \left[  \int_{0}^{\tau}dt\frac{j_{l}\left(  k_{nl}^{E}\left\vert
z\left(  t\right)  \right\vert \right)  }{k_{nl}^{E}\left\vert z\left(
t\right)  \right\vert }\cos\left[  \left(  \omega_{nl}^{E}+\omega_{0}\right)
t\right]  \right]  ^{2}+\left[  \int_{0}^{\tau}dt\frac{j_{l}\left(  k_{nl}%
^{E}\left\vert z\left(  t\right)  \right\vert \right)  }{k_{nl}^{E}\left\vert
z\left(  t\right)  \right\vert }\sin\left[  \left(  \omega_{nl}^{E}+\omega
_{0}\right)  t\right]  \right]  ^{2}\right\}  ,
\end{align}
where the superscript \textquotedblleft$Ed$\textquotedblright\ refers to the
electric dipole radiation frequency which corresponds to the natural frequency
of the oscillator,
\[
\omega_{nl}^{Ed}\approxeq\omega_{0}.
\]
\ In the large-cavity limit, this becomes from Eq. (\ref{SumC})
\begin{align}
&  \left\langle W_{l,0}\left(  \tau\right)  \right\rangle =\frac{e^{2}}%
{8M}\int_{0}^{\infty}d\omega^{E}\frac{\mathsf{R}}{\pi c}\left[  \frac
{16\pi\left(  \omega_{nl}^{Ed}\right)  ^{3}}{c^{2}\mathsf{R}}\left\langle
J_{rad}\right\rangle \right]  l\left(  l+1\right)  \left\vert Y_{l,0}%
\right\vert ^{2}\nonumber\\
&  \times\left\{  \left[  \int_{0}^{\tau}dt\frac{j_{l}\left(  k_{nl}%
^{E}\left\vert z\left(  t\right)  \right\vert \right)  }{k_{nl}^{E}\left\vert
z\left(  t\right)  \right\vert }\cos\left[  \left(  \omega_{nl}^{E}-\omega
_{0}\right)  t\right]  \right]  ^{2}+\left[  \int_{0}^{\tau}dt\frac
{j_{l}\left(  k_{nl}^{E}\left\vert z\left(  t\right)  \right\vert \right)
}{k_{nl}^{E}\left\vert z\left(  t\right)  \right\vert }\sin\left[  \left(
\omega_{nl}^{E}-\omega_{0}\right)  t\right]  \right]  ^{2}\right. \nonumber\\
&  +\left.  \left[  \int_{0}^{\tau}dt\frac{j_{l}\left(  k_{nl}^{E}\left\vert
z\left(  t\right)  \right\vert \right)  }{k_{nl}^{E}\left\vert z\left(
t\right)  \right\vert }\cos\left[  \left(  \omega_{nl}^{E}+\omega_{0}\right)
t\right]  \right]  ^{2}+\left[  \int_{0}^{\tau}dt\frac{j_{l}\left(  k_{nl}%
^{E}\left\vert z\left(  t\right)  \right\vert \right)  }{k_{nl}^{E}\left\vert
z\left(  t\right)  \right\vert }\sin\left[  \left(  \omega_{nl}^{E}+\omega
_{0}\right)  t\right]  \right]  ^{2}\right\} \nonumber\\
&  =\frac{2e^{2}}{M}\int_{0}^{\infty}d\omega^{E}\left[  \frac{\left(
\omega_{nl}^{Ed}\right)  ^{3}}{c^{3}}\left\langle J_{rad}\right\rangle
\right]  l\left(  l+1\right)  \left\vert Y_{l,0}\right\vert ^{2}\nonumber\\
&  \times\left\{  \left[  \int_{0}^{\tau}dt\frac{j_{l}\left(  k_{nl}%
^{E}\left\vert z\left(  t\right)  \right\vert \right)  }{k_{nl}^{E}\left\vert
z\left(  t\right)  \right\vert }\cos\left[  \left(  \omega_{nl}^{E}-\omega
_{0}\right)  t\right]  \right]  ^{2}+\left[  \int_{0}^{\tau}dt\frac
{j_{l}\left(  k_{nl}^{E}\left\vert z\left(  t\right)  \right\vert \right)
}{k_{nl}^{E}\left\vert z\left(  t\right)  \right\vert }\sin\left[  \left(
\omega_{nl}^{E}-\omega_{0}\right)  t\right]  \right]  ^{2}\right. \nonumber\\
&  +\left.  \left[  \int_{0}^{\tau}dt\frac{j_{l}\left(  k_{nl}^{E}\left\vert
z\left(  t\right)  \right\vert \right)  }{k_{nl}^{E}\left\vert z\left(
t\right)  \right\vert }\cos\left[  \left(  \omega_{nl}^{E}+\omega_{0}\right)
t\right]  \right]  ^{2}+\left[  \int_{0}^{\tau}dt\frac{j_{l}\left(  k_{nl}%
^{E}\left\vert z\left(  t\right)  \right\vert \right)  }{k_{nl}^{E}\left\vert
z\left(  t\right)  \right\vert }\sin\left[  \left(  \omega_{nl}^{E}+\omega
_{0}\right)  t\right]  \right]  ^{2}\right\}  . \label{Wtau10}%
\end{align}
We will use this equation (\ref{Wtau10})\ repeatedly for the average energy
gained by the charged oscillator from the random classical electromagnetic
radiation. \ 

\section{Ground State of the Charged Harmonic Oscillator in Zero-Point
Radiation}

\subsection{Small-Source Approximation}

When we are dealing with the \textit{small-source} limit, we require a small
oscillation for the charged oscillator. \ The ground state is the lowest state
for the system.\ \ Therefore, we assume that the argument of the spherical
Bessel function is small and use\cite{Jackson741}%
\begin{equation}
j_{l}\left(  x\right)  \approxeq\frac{x^{l}}{\left(  2l+1\right)  !!}.
\label{jonesmall}%
\end{equation}
Now the excursion of the oscillator is given in Eq. (\ref{ztZ0}) and in the
small-argument limit, the integrals in Eq. (\ref{Wtau10}) become%
\begin{align}
&  \int_{0}^{\tau}dt\frac{j_{l}\left(  k_{nl}^{E}Z_{0}\cos\left(  \omega
_{0}t\right)  \right)  }{k_{nl}^{E}Z_{0}\cos\left(  \omega_{0}t\right)  }%
\cos\left[  \left(  \omega_{nl}^{E}\mp\omega_{0}\right)  t\right] \nonumber\\
&  \approxeq\frac{\left(  k_{nl}^{E}\right)  ^{l-1}\left(  Z_{0}\right)
^{l-1}}{\left(  2l+1\right)  !!}\int_{0}^{\tau}dt\left[  \cos\left(
\omega_{0}t\right)  \right]  ^{l-1}\cos\left[  \left(  \omega_{nl}^{E}%
\mp\omega_{0}\right)  t\right]  \label{N2piC}%
\end{align}
and
\begin{align}
&  \int_{0}^{\tau}dt\frac{j_{l}\left(  k_{nl}^{E}Z_{0}\cos\left(  \omega
_{0}t\right)  \right)  }{k_{nl}^{E}Z_{0}\cos\left(  \omega_{0}t\right)  }%
\sin\left[  \left(  \omega_{nl}^{E}\mp\omega_{0}\right)  t\right] \nonumber\\
&  \approxeq\frac{\left(  k_{nl}^{E}\right)  ^{l-1}\left(  Z_{0}\right)
^{l-1}}{\left(  2l+1\right)  !!}\int_{0}^{\tau}dt\left[  \cos\left(
\omega_{0}t\right)  \right]  ^{l-1}\sin\left[  \left(  \omega_{nl}^{E}%
\mp\omega_{0}\right)  t\right]  . \label{N2piS}%
\end{align}
Therefore equation (\ref{Wtau10}) becomes%
\begin{align}
&  \left\langle W_{l,0}\left(  \tau\right)  \right\rangle _{\theta}\nonumber\\
&  =\frac{2e^{2}}{M}\int_{0}^{\infty}d\omega^{E}\left[  \frac{\left(
\omega^{E}\right)  ^{3}}{c^{3}}\left\langle J_{rad}\left(  \omega^{E}\right)
\right\rangle _{\theta}\right]  l\left(  l+1\right)  \left\vert Y_{l,0}%
\right\vert ^{2}\left[  \frac{\left(  k_{nl}^{E}\right)  ^{l-1}\left(
Z_{0}\right)  ^{l-1}}{\left(  2l+1\right)  !!}\right]  ^{2}\nonumber\\
&  \times\left[  \int_{0}^{\tau}dt\left[  \cos\left(  \omega_{0}t\right)
\right]  ^{l-1}\cos\left[  \left(  \omega_{nl}^{E}\mp\omega_{0}\right)
t\right]  \right]  ^{2}+\left[  \int_{0}^{\tau}dt\left[  \cos\left(
\omega_{0}t\right)  \right]  ^{l-1}\sin\left[  \left(  \omega_{nl}^{E}%
\mp\omega_{0}\right)  t\right]  \right]  ^{2}. \label{Wtau11}%
\end{align}

\subsection{Planck's Dipole Radiation Case, $l=1$}

The case $l=1$ was treated by Planck at the end of the 19th century and is
repeated in other publications.\cite{Lavenda} \ This dipole-radiation case
corresponds to ignoring the position dependence of the driving random
radiation field in Eq. (\ref{Wtau8}) and involves the integrals from Eqs.
(\ref{N2piC}) and (\ref{N2piS}) giving%
\begin{equation}
\int_{0}^{\tau}dt\cos\left[  \left(  \omega_{nl}^{E}\mp\omega_{0}\right)
t\right]  =\frac{\sin\left[  \left(  \omega_{nl}^{E}\mp\omega_{0}\right)
\tau\right]  }{\left(  \omega_{nl}^{E}\mp\omega_{0}\right)  } \label{Itcos}%
\end{equation}
and%

\begin{equation}
\int_{0}^{\tau}dt\sin\left[  \left(  \omega_{nl}^{E}\mp\omega_{0}\right)
t\right]  =\frac{1-\cos\left[  \left(  \omega_{nl}^{E}\mp\omega_{0}\right)
\tau\right]  }{\left(  \omega_{nl}^{E}\mp\omega_{0}\right)  }. \label{Itsin}%
\end{equation}
For the \textit{dipole} case, $l=1$, there is no indication of the motion of
the dipole, since $\left(  Z_{0}\right)  ^{l-1}=\left(  Z_{0}\right)  ^{0}=1.$
\ Inserting these expressions into Eq. (\ref{Wtau10}), we have%
\begin{align}
\left\langle W_{1,0}\left(  \tau\right)  \right\rangle _{\theta}  &
=\frac{2e^{2}}{M}\int_{0}^{\infty}d\omega^{E}\left[  \frac{\left(  \omega
^{E}\right)  ^{3}}{c^{3}}\left\langle \left[  J_{rad}\left(  \omega
^{E}\right)  \right]  \right\rangle _{\theta}\right]  2\left\vert \sqrt
{\frac{3}{4\pi}}\right\vert ^{2}\left[  \frac{1}{3}\right]  ^{2}\nonumber\\
&  \times\left\{  \left[  \frac{\sin\left[  \left(  \omega_{nl}^{E}\mp
\omega_{0}\right)  \tau\right]  }{\left(  \omega_{nl}^{E}\mp\omega_{0}\right)
}\right]  ^{2}+\left[  \frac{1-\cos\left[  \left(  \omega_{nl}^{E}\mp
\omega_{0}\right)  \tau\right]  }{\left(  \omega_{nl}^{E}\mp\omega_{0}\right)
}\right]  ^{2}\right\}  _{l=1,m=0}\nonumber\\
&  =\left(  \frac{3}{4\pi}\right)  \frac{2\left(  2\right)  e^{2}}{\left(
3\right)  ^{2}Mc^{3}}\int_{0}^{\infty}d\omega^{E}\left[  \left(  \omega
^{E}\right)  ^{3}\left\langle J_{rad}\left(  \omega^{E}\right)  \right\rangle
_{\theta}\right] \nonumber\\
&  \times\left\{  \left[  \frac{\sin\left[  \left(  \omega_{nl}^{E}\mp
\omega_{0}\right)  \tau\right]  }{\left(  \omega_{nl}^{E}\mp\omega_{0}\right)
}\right]  ^{2}+\left[  \frac{1-\cos\left[  \left(  \omega_{nl}^{E}\mp
\omega_{0}\right)  \tau\right]  }{\left(  \omega_{nl}^{E}\mp\omega_{0}\right)
}\right]  ^{2}\right\} \nonumber\\
&  \approxeq\frac{e^{2}}{3\pi Mc^{3}}\omega_{0}^{3}\left\langle \left[
J_{rad}\left(  \omega_{0}\right)  \right]  \right\rangle _{\theta}2\pi
\tau=\frac{2e^{2}}{3Mc^{3}}\omega_{0}^{2}\left[  U_{rad}\left(  \omega
_{0}\right)  \right]  \tau,
\end{align}
since the two functions involving $\tau$ are sharply peaked at $\omega
^{E}=\omega_{0}$. \ Note that the average radiation energy is $\left[
U_{rad}\left(  \omega_{0}\right)  \right]  =\left\langle J_{rad}\right\rangle
_{\theta}\omega_{0}$. \ 

The average power radiated in the nonrelativistic dipole approximation for the
charged linear oscillator is the familiar Larmor formula
\begin{equation}
\left\langle P_{1,0}^{loss}\right\rangle _{\theta}=\frac{2}{3}\frac{\omega
_{0}^{4}}{c^{3}}\left\langle \left[  Z_{0}\cos\left(  \omega_{0}t\right)
\right]  ^{2}\right\rangle _{\theta}=\frac{2}{3}\frac{\omega_{0}^{4}}{c^{3}%
}\left(  \frac{\left\langle J_{e}\right\rangle _{\theta}}{M\omega_{0}}\right)
=\frac{2}{3}\frac{\omega_{0}^{4}}{c^{3}}\left(  \frac{\left\langle
\mathcal{E}_{e}\left(  \omega_{0}\right)  \right\rangle _{\theta}}{M\omega
_{0}^{2}}\right)  . \label{Ploss1c0}%
\end{equation}
\ Thus we obtain Planck's result that the average energy $\left\langle \left[
\mathcal{E}\left(  \omega_{0}\right)  \right]  \right\rangle _{\theta}%
=\hbar\omega_{0}/2$ of the mechanical oscillator is the same as the average
energy $\left[  U_{rad}\left(  \omega_{0}\right)  \right]  $ for the radiation
modes at the same frequency as the mechanical frequency of the oscillator%
\begin{equation}
\left\langle \left[  \mathcal{E}_{e}\left(  \omega_{0}\right)  \right]
\right\rangle _{\theta}\mathcal{=}\left[  U_{rad}\left(  \omega_{0}\right)
\right]  \text{ \ and \ }\left\langle J_{e}\right\rangle _{\theta
}=\left\langle J_{rad}\right\rangle _{\theta}=\hbar/2.
\end{equation}

\section{Quadrupole Radiation Emission Calculation}

The quadrupole radiation emission is less familiar. The charged linear
oscillator gives rise to a vector potential%

\begin{align}
&  \mathbf{A}\left(  \mathbf{r},t\right) \nonumber\\
&  =\int d^{3}r^{\prime}\int dt^{\prime}\frac{\delta\left(  t-t^{\prime
}-\left\vert \mathbf{r-r}^{\prime}\right\vert /c\right)  }{\left\vert
\mathbf{r-r}^{\prime}\right\vert }\frac{\mathbf{J(r}^{\prime},t^{\prime})}%
{c}\nonumber\\
&  =\int d^{3}r^{\prime}\int dt^{\prime}\frac{\delta\left(  t-t^{\prime
}-\left\vert \mathbf{r-r}^{\prime}\right\vert /c\right)  }{\left\vert
\mathbf{r-r}^{\prime}\right\vert }\frac{e\widehat{z}Z_{0}\left(  -\omega
_{0}\right)  \sin\left(  \omega_{0}t^{\prime}\right)  \delta^{3}\left[
\mathbf{r}^{\prime}-\widehat{z}Z_{0}\cos\left(  \omega_{0}t^{\prime}\right)
\right]  }{c}\nonumber\\
&  =-\int d^{3}r^{\prime}\int dt^{\prime}\frac{\delta\left(  t-t^{\prime
}-\left\vert \mathbf{r-r}^{\prime}\right\vert /c\right)  }{\left\vert
\mathbf{r-r}^{\prime}\right\vert }\frac{e\widehat{z}Z_{0}\omega_{0}\sin\left(
\omega_{0}t^{\prime}\right)  }{c}\left[  \delta^{3}\left(  \mathbf{r}^{\prime
}\right)  \right. \nonumber\\
&  \left.  -Z_{0}\cos\left(  \omega_{0}t^{\prime}\right)  \widehat{z}%
\cdot\nabla^{\prime}\delta^{3}\left(  \mathbf{r}^{\prime}\right)
+....\right]
\end{align}
with an infinite number of terms. \ The term in $\delta^{3}\left(
\mathbf{r}\right)  $ corresponds to the dipole term which is given in all the
textbooks. \ \ At present, we are interested in just the quadrupole term
$\mathbf{A}^{\left(  2\right)  }\left(  \mathbf{r},t\right)  $. \ Now we
integrate by parts on $\nabla^{\prime}\delta^{3}\left(  \mathbf{r}^{\prime
}\right)  $ so as move the gradient outside and to take advantage of the delta
function. \ We find
\begin{equation}
\mathbf{A}^{\left(  2\right)  }\left(  \mathbf{r},t\right)  =\widehat{z}%
\frac{eZ_{0}^{2}\omega_{0}}{2}\widehat{z}\cdot\nabla\left[  \frac{\sin\left[
2\omega_{0}\left(  t-r/c\right)  \right]  }{cr}\right]  .
\end{equation}
We are interested in only the emitted radiation, and obtain%
\begin{equation}
\mathbf{A}^{\left(  2\right)  }\left(  \mathbf{r},t\right)  \approxeq
-\widehat{z}\frac{eZ_{0}^{2}\omega_{0}}{2cr}\left(  \frac{2\omega_{0}}%
{c}\right)  \left(  \cos\theta\right)  \left\{  \cos\left[  2\omega_{0}\left(
ct-r\right)  /c\right]  \right\}  .
\end{equation}
Using $\mathbf{B=}\nabla\mathbf{\times A,}$ we find in the radiation zone,%
\begin{equation}
\mathbf{B(r},t)\approxeq\left[  \widehat{\phi}\frac{eZ_{0}^{2}\omega_{0}}%
{2c}\left(  \frac{2\omega_{0}}{c}\right)  ^{2}\left(  \cos\theta\sin
\theta\right)  \right]  \frac{1}{r}\left[  \sin\left[  2\omega_{0}\left(
ct-r\right)  /c\right]  \right]
\end{equation}
We now calculate the power radiated per unit solid angle from $dP_{quad}%
^{loss}/d\Omega=r^{2}\left[  e/\left(  4\pi\right)  \right]  \widehat{r}%
\cdot\left(  \mathbf{E\times H}\right)  \mathbf{=}\left[  e/\left(
4\pi\right)  \right]  r^{2}B^{2}$ and integrate over all solid angles to
obtain $\int_{0}^{2\pi}d\phi\int_{0}^{\pi}d\theta\sin\theta\left(  \cos
\theta\sin\theta\right)  ^{2}=8\pi/15.$ \ Thus, using Eq. (\ref{Z0}), we
obtain the total time-average power emitted as quadrupole radiation as%
\begin{align}
\left\langle P_{quad}^{loss}\right\rangle _{t}  &  =\frac{c}{4\pi}\left[
\frac{eZ_{0}^{2}\omega_{0}}{2c}\left(  \frac{2\omega_{0}}{c}\right)
^{2}\right]  ^{2}\frac{8\pi}{15}\left\langle \left[  \sin\left[  2\omega
_{0}\left(  ct-r\right)  /c\right]  \right]  ^{2}\right\rangle _{t}\nonumber\\
&  =\frac{16e^{2}\omega_{0}^{4}}{\left(  15\right)  c^{5}M^{2}}J_{e}^{2}%
=\frac{8}{5}\left[  \left(  \frac{2e^{2}}{3Mc^{3}}\right)  \omega_{0}%
^{2}\right]  J_{e}\omega_{0}\left(  \frac{J_{e}\omega_{0}}{Mc^{2}}\right)  .
\label{Plossquad}%
\end{align}
Notice the dependence on $1/c^{5}$ whereas the dipole radiation emission is
$1/c^{3}.$

\subsection{Quadrupole Radiation Case, $l=2$}

We wish to compare this radiation loss with the radiation gained from
zero-point radiation. \ For the case $l=2$, the integrals in Eqs.
(\ref{N2piC}) and (\ref{N2piS}) become for the resonant terms
\begin{align}
&  \int_{0}^{\tau}dt\left[  \cos\left(  \omega_{0}t\right)  \right]
\cos\left[  \left(  \omega_{nl}^{E}-\omega_{0}\right)  t\right] \nonumber\\
&  =\int_{0}^{\tau}dt\frac{1}{2}\left\{  \cos\left[  \left(  \omega
^{E}-2\omega_{0}\right)  t\right]  +\cos\left[  \omega^{E}t\right]  \right\}
\nonumber\\
&  =\frac{1}{2}\left\{  \frac{\sin\left[  \left(  \omega^{E}-2\omega
_{0}\right)  \tau\right]  }{\left[  \left(  \omega^{E}-2\omega_{0}\right)
\right]  }+\frac{\sin\left[  \omega^{E}\tau\right]  }{\omega^{E}}\right\}
\end{align}
and
\begin{align}
&  \int_{0}^{\tau}dt\left[  \cos\left(  \omega_{0}t\right)  \right]
\sin\left[  \left(  \omega_{nl}^{E}-\omega_{0}\right)  t\right] \nonumber\\
&  =\int_{0}^{\tau}dt\frac{1}{2}\left\{  \sin\left[  \left(  \omega
^{E}-2\omega_{0}\right)  t\right]  +\sin\left[  \omega^{E}t\right]  \right\}
\nonumber\\
&  =\frac{1}{2}\left\{  \frac{1-\cos\left[  \left(  \omega^{E}-2\omega
_{0}\right)  \tau\right]  }{\left[  \left(  \omega^{E}-2\omega_{0}\right)
\right]  }+\frac{1-\cos\left[  \omega^{E}\tau\right]  }{\omega^{E}}\right\}  .
\end{align}
Only the terms with the quantity $\left(  \omega^{E}-2\omega_{0}\right)  $ in
the denominator become large when $\omega^{E}\approxeq2\omega_{0}$. \ The
random classical zero-point radiation is independently correlated for each
normal mode so that the radiation at $2\omega_{0}$ is not correlated with that
at $\omega_{0}$. \ Thus the energy gain equation in (\ref{Wtau10}) becomes for
$l=2$%

\begin{align}
&  \left\langle W_{2,0}\left(  \tau\right)  \right\rangle _{\theta}\nonumber\\
&  =\frac{2e^{2}}{M}\int_{0}^{\infty}d\omega^{E}\left[  \frac{\left(
\omega_{0}\right)  ^{3}}{c^{3}}\left\langle J_{rad}\right\rangle _{\theta
}\right]  6\left[  \frac{5}{4\pi}\right]  \left[  \frac{\left(  k_{nl}%
^{E}\right)  ^{1}\left(  Z_{0}\right)  ^{1}}{15}\right]  ^{2}\nonumber\\
&  \times\left[  \int_{0}^{\tau}dt\left[  \cos\left(  \omega_{0}t\right)
\right]  ^{1}\cos\left[  \left(  \omega_{nl}^{E}\mp\omega_{0}\right)
t\right]  \right]  ^{2}+\left[  \int_{0}^{\tau}dt\left[  \cos\left(
\omega_{0}t\right)  \right]  ^{1}\sin\left[  \left(  \omega_{nl}^{E}\mp
\omega_{0}\right)  t\right]  \right]  ^{2}\nonumber\\
&  =\frac{2\left(  2\right)  3e^{2}}{\left(  15\right)  ^{2}M}\left(  \frac
{5}{4\pi}\right)  \int_{0}^{\infty}d\omega^{E}\left[  \frac{\left(  \omega
^{E}\right)  ^{3}}{c^{3}}\left\langle \left[  J_{rad}\left(  \omega
^{E}\right)  \right]  \right\rangle _{\theta}\right]  \left(  k_{nl}%
^{E}\right)  ^{2}\left(  \frac{2J_{e}}{M\omega_{0}}\right) \nonumber\\
&  \times\left\{  \left[  \frac{\sin\left[  \left(  \omega^{E}-2\omega
_{0}\right)  \tau\right]  }{\left[  \left(  \omega^{E}-2\omega_{0}\right)
\right]  }+\frac{\sin\left[  \omega^{E}\tau\right]  }{\omega^{E}}\right]
^{2}+\left[  \frac{1-\cos\left[  \left(  \omega^{E}-2\omega_{0}\right)
\tau\right]  }{\left[  \left(  \omega^{E}-2\omega_{0}\right)  \right]  }%
+\frac{1-\cos\left[  \omega^{E}\tau\right]  }{\omega^{E}}\right]  ^{2}\right\}
\nonumber\\
&  =\frac{8}{5}\left(  \frac{2e^{2}}{3Mc^{3}}\right)  \left[  \left(
\omega_{0}\right)  ^{2}\left\langle \left[  J_{rad}\left(  2\omega_{0}\right)
\right]  \right\rangle _{\theta}\omega_{0}\right]  \left(  \frac{J_{e}%
\omega_{0}}{Mc^{2}}\right)  \tau\label{W20tau}%
\end{align}
where we have used
\begin{equation}
Y_{2,0}\left(  \theta=0\right)  =\sqrt{\frac{5}{4\pi}}\left(  \frac{3}{2}%
\cos^{2}\theta-\frac{1}{2}\right)  _{\theta=0}=\sqrt{\frac{5}{4\pi}}%
\end{equation}
and the integrals in%
\begin{equation}
\int_{-\infty}^{\infty}dx\frac{\sin^{2}\left(  xv\right)  }{x^{2}}=\pi v\text{
\ and }\int_{-\infty}^{\infty}dx\frac{\left[  1-\cos\left(  xv\right)
\right]  ^{2}}{x^{2}}=\pi v.\text{\ }%
\end{equation}
Also, it is important to note that the \textit{amplitude} of the
\textit{oscillator} motion is determined mainly by the radiation forcing at
the \textit{dipole} interaction frequency $\omega_{0}$, not at the quadrupole
frequency $2\omega_{0}$. \ The quadrupole radiation for both emission and
absorption involves higher factors of $1/c^{2}$ and so makes a much smaller
contribution than the oscillator dipole amplitude, though the stochastic
process for the oscillator is the same. \ 

We now need to compare this power loss with the quadrupole energy emitted as
radiation by the charged oscillator. \ This result in Eq. (\ref{Plossquad})
agrees with the radiation gain calculated in Eq. (\ref{W20tau}). \ Thus, for
the same oscillator motion, we find an energy balance for both dipole and
quadrupole radiation in classical zero-point radiation. \ We suggest that the
balance holds for all multipole radiation terms in the large-mass-$M$ limit. \ 

\section{Resonant Excited States}

\subsection{Preliminary Ideas}

\subsubsection{Suggestive Work by Huang and Batelaan and by Cole}

In 2015, Huang and Batelaan\cite{Batelaan} showed that a classical charged
harmonic oscillator in classical zero-point radiation would absorb radiation
energy from a transient electromagnetic pulse. \ The absorption was at
\textit{integer} multiples $n\omega_{0}$ of the the natural mechanical
oscillation frequency $\omega_{0}$. \ Furthermore, in 2018, Cole\cite{Cole}
pointed out that for a charged particle in a Coulomb potential, there were
large resonances for driving radiation at \textit{integer} multiples
$n\omega_{e}$ of the mechanical frequency $\omega_{e}$ of orbital motion.
\ There were also resonant eccentricities of the orbital motion. \ Cole's
resonances corresponded to absorption of both energy and angular momentum by
the charged particle sufficient to balance the loss of mechanical energy and
angular momentum due to radiation emission. \ However, Cole did not consider
driving by zero-point radiation, but rather treated driving by a
\textit{circularly polarized plane wave} incident normal to the orbital plane
of the charged particle in the Coulomb potential, and of various wave
amplitudes. \ This earlier work suggests that, for a charged mechanical
oscillator in electromagnetic radiation, one might look for resonances in the
spherical multipole radiation at integer multiples of the mechanical
oscillation frequency. \ 

\subsubsection{Resonance for Electromagnetic Waves}

A mechanical oscillator has exactly one resonant frequency, that of its
natural frequency of oscillation $\omega_{0}$. \ Thus when pushing a swing,
there is only one resonant frequency. \ However, electromagnetic radiation is
not limited to the fundamental oscillator frequency\textit{ }$\omega_{0}$ for
the charged harmonic oscillator because of the \textit{position-dependence of
the driving electric field} $\mathbf{E}\left(  \mathbf{r},t\right)
=\mathbf{E}\left[  \widehat{z}z\left(  t\right)  ,t\right]  $. \ This position
dependence is crucial for the excited resonant states. \ Classical zero-point
radiation provides driving forces at all frequencies and all spherical
multipoles, as indicated in Eq. (\ref{Ezprt}). \ The presence of the
$\widehat{z}z\left(  t\right)  $ in the argument of the radiation field
$\mathbf{E}\left[  \widehat{z}z\left(  t\right)  ,t\right]  $ means that the
oscillator equation of motion corresponds to a \textit{parametric oscillator}
with an entirely new set of resonances at integer multiples of its natural
oscillation frequency $\omega_{0}$. \ The position-dependence of the driving
electromagnetic field is an aspect which sometimes seems overlooked when
considering resonant excited states.

\subsubsection{Dipole Behavior for Resonant Excited States}

In the small-source-large-$c$ approximation appropriate for
charged-linear-oscillator mechanical motion, the radiation from the spherical
multipole moments of order $l$ goes as $1/c^{2l+1}$. \ For example, the dipole
radiation involves power radiated at $1/c^{3}$ whereas \ quadrupole radiation
power is at $1/c^{5}$. \ Indeed, when the mechanical speed is small, the
lowest possible value of $l$ always dominates the radiation energy loss due to
emission. \ For a point charge, the radiation power going as $1/c^{3}$ always
dominates. \ 

\textit{Numerical} integration of the last lines in Eq. (\ref{Wtau10})
regarded as functions of the zero-point driving frequency $\omega^{E}$
suggests that equally-spaced peaks in the mechanical motion appear when
$\omega^{E}=m\omega_{0}$ for odd-integral $m,$ and the peaks (beyond the
first) all involve about the same height and width in $\omega^{E}$.
\ Accordingly, we will look for resonant excited states involving dipole
oscillator emission when the radiation absorption is close to an odd integer
multiple of $\omega_{0}$. \ 

\subsection{The Bessel Function $j_{1}$\ for Dipole Radiation}

\subsubsection{Consideration when the Argument of the Spherical Bessel
Function Vanishes}

Although for the ground state in Eq. (\ref{jonesmall}), we considered only the
small-argument limit for $j_{1}\left(  x\right)  $, we cannot do this for the
excited states; rather, we will need the full expression. \ We can express all
the \textit{spherical} Bessel functions in terms of sine and cosine functions
and inverse powers of the argument. \ For example, when $l=1$, the spherical
Bessel function is
\begin{equation}
j_{1}\left(  x\right)  =\frac{\sin\left(  x\right)  }{x^{2}}-\frac{\cos\left(
x\right)  }{x}. \label{j1x}%
\end{equation}
Now in Eq. (\ref{j1x}), the inverse powers in $x$ suggest singularities when
the argument $x$ vanishes. \ However, all the spherical Bessel functions are
finite for all real values of the argument $x$, even at $x=0$. \ Indeed,
expanding the sine and cosine function about $x=0$ shows that the apparent
singularity cancels, and we have for $j_{1}\left(  x\right)  /x$ for small
argument $x$%
\begin{align}
\frac{j_{1}\left(  x\right)  }{x}  &  =\frac{\sin\left(  x\right)  }{x^{3}%
}-\frac{\cos\left(  x\right)  }{x^{2}}\nonumber\\
&  =\frac{1}{x^{3}}\left[  x-\frac{x^{3}}{3!}+\frac{x^{5}}{5!}-...\right]
-\frac{1}{x^{2}}\left[  1-\frac{x^{2}}{2!}+\frac{x^{4}}{4!}-...\right]
\nonumber\\
&  =\frac{1}{3}-\frac{x^{2}}{30}+...
\end{align}

Moreover, there are additional considerations. The sine and cosine functions
appearing in Eq. (\ref{j1x}) are periodic with sign changes for a period $\pi
$. \ Furthermore, the function of interest in Eq. (\ref{Wtau10}) involves
$j_{1}\left(  k^{E}Z_{0}\cos\left(  \omega_{0}t\right)  \right)  /\left[
k^{E}Z_{0}\cos\left(  \omega_{0}t\right)  \right]  $. \ This function is
periodic in time $t$ because of the argument $k^{E}Z_{0}\cos\left(  \omega
_{0}t\right)  $. \ The Bessel function $j_{1}\left(  x\right)  $ is an odd
function of $x,$ whereas $x$ is also an odd function of $x\,$. \ Therefore the
function $j_{1}\left(  k^{E}Z_{0}\cos\left(  \omega_{0}t\right)  \right)
/\left[  k^{E}Z_{0}\cos\left(  \omega_{0}t\right)  \right]  $ is an even
function of its argument,\ and has only \textit{even} powers of the argument
$k^{E}Z_{0}\cos\left(  \omega_{0}t\right)  $. \ The lead term in $j_{1}\left(
k^{E}Z_{0}\cos\left(  \omega_{0}t\right)  \right)  /\left[  k^{E}Z_{0}%
\cos\left(  \omega_{0}t\right)  \right]  $ is a constant, giving a maximum
whenever $\cos\left(  \omega_{0}t\right)  $ vanishes, namely at $\omega
_{0}t=m\pi+\pi/2$ where $m$ is an integer. \ The function $j_{1}\left(
k^{E}Z_{0}\cos\left(  \omega_{0}t\right)  \right)  /\left[  k^{E}Z_{0}%
\cos\left(  \omega_{0}t\right)  \right]  $ is a smooth continuous function of
$t$, even when $\cos\left(  \omega_{0}t\right)  $ goes to zero. \ Thus\ for
$\omega_{0}t=m\pi+\pi/2+x$ where the absolute value $\left\vert x\right\vert $
is a small quantity less than $\pi$, the spherical Bessel function becomes
when $k^{E}Z_{0}x$ is small,
\begin{align}
\frac{j_{1}(k^{E}Z_{0}\cos(\omega_{0}t))}{k^{E}Z_{0}\cos(\omega_{0}t)}  &
=\frac{j_{1}(k^{E}Z_{0}\left\vert \cos(\omega_{0}t)\right\vert )}{k^{E}%
Z_{0}\left\vert \cos(\omega_{0}t)\right\vert }=\frac{j_{1}(k^{E}%
Z_{0}\left\vert \cos\left(  m\pi+\pi/2+x\right)  \right\vert )}{k^{E}%
Z_{0}\left\vert \cos(m\pi+\pi/2+x)\right\vert }\nonumber\\
&  =\left(  \frac{j_{1}(k^{E}Z_{0}\sin\left(  x\right)  )}{k^{E}Z_{0}%
\sin\left(  x\right)  }\right) \\
&  =\left(  \frac{\sin\left(  k^{E}Z_{0}\sin\left(  x\right)  \right)
}{\left[  k^{E}Z_{0}\sin\left(  x\right)  \right]  ^{3}}-\frac{\cos\left(
k^{E}Z_{0}\sin\left(  x\right)  \right)  }{\left[  k^{E}Z_{0}\sin\left(
x\right)  \right]  ^{2}}\right) \nonumber\\
&  =\left(  \frac{1}{3}-\frac{\left[  k^{E}Z_{0}\sin\left(  x\right)  \right]
^{2}}{30}+~...\right)  \approxeq\frac{1}{3}-\frac{\left[  k^{E}Z_{0}x\right]
^{2}}{30}+... \label{sinA}%
\end{align}
Note that here $x=\omega_{0}t-\left(  m\pi+\pi/2\right)  =\omega_{0}\left(
t-t_{\ast}\right)  $ is a small displacement from $\omega_{0}t_{\ast}=m\pi
+\pi/2$. \ This corresponds to a peak for the absolute value $\left\vert
j_{1}\left(  k^{E}Z_{0}\cos\left(  \omega_{0}t\right)  \right)  /\left[
k^{E}Z_{0}\cos\left(  \omega_{0}t\right)  \right]  \right\vert $ at
$\omega_{0}t=\omega_{0}t_{\ast}=m\pi+\pi/2$ for any integer $m\geq1$. \ At the
peak, the cosine vanishes, $\cos\left(  \omega_{0}t_{\ast}\right)
=\cos\left(  \pi/2\right)  =0$, the peak has value $1/3$, and then the
function of $t$ decreases in absolute magnitude on either side of the peak. \ 

\subsubsection{Fourier Time Series in Zero-Point Radiation}

Now the time integral in Eq. (\ref{Wtau10}) involves many oscillations of the
charged oscillator. \ However, the function $\left[  j_{1}(k^{E}Z_{0}%
\cos(\omega_{0}t))\right]  /(k^{E}Z_{0}\cos(\omega_{0}t))$ in Eq.
(\ref{sinA})\ is a real periodic function of time $t$ with period $2\pi
/\omega_{0}$. \ Our calculation in Eq. (\ref{sinA}) shows that all of the
maxima of the function at time $t=\left(  m\pi+\pi/2+x\right)  /\omega_{0}$
have the height approximately $1/3$ for $m\geq1$. \ 

Now if the time $\tau$ is large compared to the the periods $2\pi/\omega_{0}$,
then $\omega_{0}\tau\approxeq p2\pi$ where $p$ is an integer, and the needed
integral is $p$ times the integral over a single period,
\begin{equation}
\int_{0}^{\tau}dt\frac{j_{1}(k^{E}Z_{0}\cos(\omega_{0}t))}{k^{E}Z_{0}%
\cos(\omega_{0}t)}\cos\left[  \left(  \omega^{E}-\omega_{0}\right)  t\right]
\approxeq p\int_{0}^{2\pi/\omega_{0}}dt\frac{j_{1}(k^{E}Z_{0}\left\vert
\cos(\omega_{0}t)\right\vert )}{k^{E}Z_{0}\left\vert \cos(\omega
_{0}t)\right\vert }\cos\left[  \left(  \omega^{E}-\omega_{0}\right)  t\right]
. \label{idtj1}%
\end{equation}
The function $j_{1}\left(  k^{E}Z_{0}\left\vert \cos\left(  \omega
_{0}t\right)  \right\vert \right)  /\left[  k^{E}Z_{0}\left\vert \cos\left(
\omega_{0}t\right)  \right\vert \right]  $ is clearly even both about
$\omega_{0}t=\pi$ since $\left\vert \cos(\pi-\omega_{0}t)\right\vert
=\left\vert \cos\left(  \pi+\omega_{0}t\right)  \right\vert =\left\vert
\pm\cos\left(  \omega_{0}t\right)  \right\vert $ and about $\omega_{0}%
t=\pi/2,$ since $\left\vert \cos(\pi/2-\omega_{0}t)\right\vert =\left\vert
\cos\left(  \pi/2+\omega_{0}t\right)  \right\vert =\left\vert \pm\sin\left(
\omega_{0}t\right)  \right\vert $ But then if the ratio $\omega^{E}/\omega
_{0}$ is an even integer, the cosine function $\cos\left[  \left(  \omega
^{E}-\omega_{0}\right)  t\right]  $ in Eq. (\ref{idtj1}) is odd about
$\omega_{0}t=\pm\pi/2,$ and the integral in Eq. (\ref{idtj1}) will vanish.
\ Thus, the resonant excited states will involve $\omega^{E}/\omega_{0}$ as
odd, $\omega^{E}=\left(  2n+1\right)  \omega_{0}$. \ \ At resonance for the
charged mechanical system, the resonant driving frequency for the zero-point
radiation is exactly the odd integers. \ 

\subsection{Examples}

\subsubsection{Case of the Ground State, $\omega^{E}=\omega_{0}$ \ }

The ground state corresponds to the smallest amplitude $Z_{0}$ of oscillation.
\ This is a special case where the argument $k^{E}Z_{0}\cos\left(  \omega
_{0}t\right)  $ of the spherical Bessel function $j_{1}$ never gets near its
first zero. Then $j_{1}\left(  k^{E}Z_{0}\cos\left(  \omega_{0}t\right)
\right)  /\left[  k^{E}Z_{0}\cos\left(  \omega_{0}t\right)  \right]  $
$\approxeq1/3$ is simply a positive constant, and the time integration gives%
\begin{align}
&  \int_{0}^{\tau}dt\frac{j_{1}(k^{E}Z_{0}\cos(\omega_{0}t))}{k^{E}Z_{0}%
\cos(\omega_{0}t)}\cos\left[  \left(  \omega^{E}-\omega_{0}\right)  t\right]
\nonumber\\
&  \approxeq\int_{0}^{\tau}dt\frac{1}{3}\cos\left[  \left(  \omega^{E}%
-\omega_{0}\right)  t\right]  =\frac{1}{3}\left\{  \frac{\sin\left[  \left(
\omega^{E}-\omega_{0}\right)  \tau\right]  }{\omega^{E}-\omega_{0}}\right\}  ,
\end{align}
with the integral corresponding to Eq. (\ref{Itcos}) and an analogous integral
to Eq. (\ref{Itsin}) for the sine function. \ Then we have the integrals
\begin{align}
&  \int_{0}^{\infty}d\omega^{E}\left(  \omega^{E}\right)  ^{3}\left[  \int%
_{0}^{\tau}dt\frac{j_{1}(k^{E}Z_{0}\cos(\omega_{0}t))}{k^{E}Z_{0}\cos
(\omega_{0}t)}\cos\left[  \left(  \omega^{E}-\omega_{0}\right)  t\right]
\right]  ^{2}\nonumber\\
&  +\int_{0}^{\infty}d\omega^{E}\left(  \omega^{E}\right)  ^{3}\left[
\int_{0}^{\tau}dt\frac{j_{1}(k^{E}Z_{0}\cos(\omega_{0}t))}{k^{E}Z_{0}%
\cos(\omega_{0}t)}\sin\left[  \left(  \omega^{E}-\omega_{0}\right)  t\right]
\right]  ^{2}\\
&  =\int_{0}^{\infty}d\omega^{E}\left(  \omega^{E}\right)  ^{3}\left\{
\left[  \int_{0}^{\tau}dt\frac{1}{3}\cos\left[  \left(  \omega^{E}-\omega
_{0}\right)  t\right]  \right]  ^{2}+\left[  \int_{0}^{\tau}dt\frac{1}{3}\sum
in\left[  \left(  \omega^{E}-\omega_{0}\right)  t\right]  \right]
^{2}\right\} \nonumber\\
&  =\left(  \omega\,_{0}\right)  ^{3}\left(  \frac{1}{3}\right)  ^{2}2\pi
\tau.\nonumber
\end{align}
Comparing this result with our work on for the ground state, we see that that
there is agreement. \ 

The ground state is a special case involving both the $\cos\left[  \left(
\omega_{nl}^{E}-\omega_{0}\right)  t\right]  $ and the $\sin\left[  \left(
\omega_{nl}^{E}-\omega_{0}\right)  t\right]  $ functions in Eq. (\ref{Wtau10}%
), and so is larger by a factor of approximately $2$ than the terms with
$\omega^{E}\approxeq(2n+1)\omega_{0}$ for $n>0$. \ 

\subsubsection{Case of the First Possible Overtone, $\omega^{E}\approxeq
2\omega_{0}$}

This situation where the resonant frequency for the driving radiation is an
\textit{even} integer $\omega^{^{\prime}E}=2\omega_{0}$ corresponds to the
first possible resonant excited state. \ Now at the possible resonance, the
difference is $\omega^{E}-\omega_{0}=\omega_{0},$ then we have%

\begin{align}
&  \int_{0}^{\tau}dt\frac{j_{1}(k^{E}Z_{1}\cos(\omega_{0}t))}{k^{E}Z_{1}%
\cos(\omega_{0}t)}\cos\left[  \left(  \omega^{E}-\omega_{0}\right)  t\right]
\approxeq p\int_{0}^{2\pi/\omega_{0}}dt\frac{j_{1}(k^{E}Z_{1}\cos(\omega
_{0}t))}{k^{E}Z_{1}\cos(\omega_{0}t)}\cos\left[  \omega_{0}t\right]
\nonumber\\
&  \approxeq2p\int_{0}^{\pi/\omega_{0}}dt\left(  \frac{j_{1}(k^{E}%
Z_{1}\left\vert \cos(\omega_{0}t)\right\vert )}{k^{E}Z_{1}\left\vert
\cos(\omega_{0}t)\right\vert }\right)  \cos\left[  \omega_{0}t\right]  =0,
\end{align}
since $\cos\left[  \omega_{0}t\right]  $ is an even function about $\omega
_{0}t=\pi$, $\,$but is odd about $\omega_{0}t=\pi/2$. Note the
\textit{changes} in limits on the integrals. \ 

\subsubsection{Case of the First Resonant Excited State, $\omega^{E}%
\approxeq3\omega_{0}$}

On the other hand when $\omega^{E}=3\omega_{0}$, then $\cos\left[  \left(
3-1\right)  \omega_{0}t\right]  =\cos\left[  2\omega_{0}t\right]  $ which
reaches its maximum value at $\omega_{0}t=\pi$, and simply repeats the
behavior found at the beginning at $t=0$. \ Thus the integrand gives a
positive integral. \ 

\subsubsection{\textit{Resonant Excited States at }$\omega_{n}^{E}%
\approxeq\left(  2n+1\right)  \omega\,_{0}$, $n=0,1,2,...$}

It is clear that the resonant excited state $n=2$ will be driven by a
radiation frequency $\omega_{2}^{E}=5\omega_{0}$, etc. \ Reviewing the
situation, the ground state with $n=0$ is resonant at the charged oscillator's
natural frequency $\omega_{rad-0}\approxeq$ $\omega_{0}$. \ The first resonant
excited state $n=1$ is resonant at $\omega_{rad-1}=$ $3\omega_{0}$. \ For
$n=2,$ the resonance is with $\omega_{rad-2}=5\omega_{0}$. etc. \ The resonant
excited states will occur at the radiation driving frequency $\omega
_{rad-n}=\omega^{E}=\left(  2n+1\right)  \omega_{0}$. \ \ This behavior is
consistent with $SO\left(  2\right)  $ symmetry, and the resonant excited
states are at odd values of the index $n$ labeling the representations of
$SO\left(  2\right)  .$

\subsubsection{Oscillator Energies}

Now the oscillator amplitude $Z_{n}$ in the $n$th resonant excited state must
be small in order to satisfy the approximately-relativistic restriction
$v<<c$. \ However, the amplitude of the oscillation is obtained from
$\left\langle \left[  Z_{n}\cos\left(  \omega_{0}t\right)  \right]
^{2}\right\rangle _{t}=\left(  Z_{n}\right)  ^{2}/2$, where from Eq.
(\ref{Z0}), the amplitude is a fixed number
\begin{align}
Z_{n}^{2}  &  =\frac{2\left\langle J_{e-n}\right\rangle }{M\omega_{0}%
}=2\left(  2n+1\right)  \frac{\left\langle J_{e-0}\right\rangle }{M\omega_{0}%
}=2\left(  2n+1\right)  \frac{\left\langle J_{rad}\right\rangle }{M\omega_{0}%
}\nonumber\\
&  =2\left(  2n+1\right)  \frac{\hbar}{2}\frac{1}{M\omega_{0}}=\left(
2n+1\right)  \frac{\hbar}{M\omega_{0}}.
\end{align}
Accordingly, the average value of the action variable for the oscillator is
\begin{equation}
\left\langle J_{e-n}\right\rangle =\left(  2n+1\right)  \frac{\hbar}%
{2}=(n+\frac{1}{2})\hbar.
\end{equation}
The mass $M$ can be chosen arbitrary large while keeping $J_{e-n}$ constant.
\ Thus in the resonant excited states, the stochastic energy of the radiation
mode $U_{rad}$ (which is balancing the loss of energy $\mathcal{E}_{e-n}$ by
the oscillator) is the same as the stochastic energy $\mathcal{E}_{e-n}$ as
the average energy of the oscillator itself,
\begin{equation}
J_{e-n}\omega_{0}=\left[  \left(  2n+1\right)  J_{rad}\right]  \omega
_{0}=J_{rad}\left[  \left(  2n+1\right)  \omega_{0}\right]  =J_{rad}%
\omega_{rad},
\end{equation}
and the average energy matches%
\[
\left\langle J_{e-n}\omega_{0}\right\rangle =\left[  \left(  2n+1\right)
\left\langle J_{rad}\right\rangle \right]  \omega_{0}=\left\langle
J_{rad}\right\rangle \left[  \left(  2n+1\right)  \omega_{0}\right]
=\left\langle J_{rad}\right\rangle \omega_{rad}=\left(  n+\frac{1}{2}\right)
\hbar.
\]

All the frequencies are present in classical zero-point radiation, and all try
to deliver energy to the charged particle. \ However, \textit{resonance} will
occur only if the charged particle goes around the same orbit repeatedly, and
this will occur only provided that the particle's power gain from the
zero-point radiation is in balance with the power lost due to radiation
emission. \ 

\section{Energy Transitions and Bohr's Rule}

Even when in a resonant excited state, the loss of energy by the oscillator is
mainly at the \textit{dipole} frequency $\omega_{0}$, but the energy gain is
at a higher frequency $\omega_{rad-n}=\left(  2n+1\right)  \omega_{0}$.
\ There is no imbalance in the oscillator's \textit{dipole} energy during the
time when the oscillator is in the resonant excited state. \ However, the
\textit{higher} driving multipoles are, in general, \textit{not} in energy
balance unless the oscillator is in its ground state. \ The transition from
one value of $n$ to another is associated with a change in the amplitude of
the charged mechanical oscillator and also in the frequency of the
\textit{driving} radiation $\omega_{rad-n}$. \ If the integer $n$ changes by
one unit from $n$ to $n-1$, then the change in average energy of the charged
mechanical oscillator is%
\begin{equation}
\Delta U_{e=n\rightarrow n}=\left[  n+\frac{1}{2}\right]  \hbar\omega
_{0}-\left[  \left(  n-1\right)  +\frac{1}{2}\right]  \hbar\omega_{0}%
=\hbar\omega_{0},
\end{equation}
and the frequency of the transition is $\omega_{0}$ which is exactly the
natural oscillation frequency $\omega_{0}$ of the charged oscillator in empty
space. \ 

The difference in \textit{energies} of the driving radiation matches the
change in mechanical energy of the oscillator. \ In transitions between
different values of $n$, the \textit{change} in energy for the oscillator is
the same as the \textit{change} in energy for the driving radiation. In the
situation of oscillator energy \textit{balance} at each resonance excited
state labeled by $n$, it may appear as though the charged particle were not
radiating at all, since the oscillator's energy does not change. \ Net
radiation appears only on \textit{changes} of the index $n.$ \ During the
transition, the charged particle radiation is not balanced by the driving
zero-point radiation. \ The energy change satisfies Bohr's relation for the
oscillator $\Delta U_{e-i\rightarrow f}\mathcal{=}U_{e-i}-U_{e-f}=h\omega_{0}%
$. \ Once again, just as for the ground state, the stabilizing role of the
classical zero-point radiation in resonant excited states may seem completely
hidden. \ 

\section{Stochastic Processes versus Eigenvalues}

In our classical electrodynamic analysis, the random classical zero-point
radiation is a stochastic process for each normal mode of frequency $\omega$,
$H\left(  J_{rad},\omega\right)  =J_{rad}\omega.$ \ Thus, since the frequency
$\omega$ is fixed for the normal mode while the action variable $J_{rad}$ is a
stochastic process, the energy $H\left(  J_{rad},\omega\right)  $ for each
normal mode must also be a stochastic process. \ The average value for
$J_{rad}$ is $\left\langle J_{rad}\right\rangle =\hbar/2,$ independent of the
frequency $\omega.$ \ This stochastic process is then transferred to the
charged \textit{mechanical oscillator} with Hamiltonian $H(J_{e-n},\omega
_{0})=J_{e-n}\omega_{0}$ and average energy $\left\langle H(J_{e-n},\omega
_{0})\right\rangle =\left\langle J_{e-n}\right\rangle \omega_{0}=\omega
_{0}\hbar/2$. \ 

This classical description is in contrast with the quantum viewpoint which
regards the oscillator energy $\widehat{H}\left(  \omega_{0}\right)  $ for any
oscillator as an eigenvalue with no dispersion, $\left\langle \widehat{H}%
\left(  \omega_{0}\right)  \right\rangle =\hbar\omega_{0}/2$. \ The average
value of the classical analysis agrees with the expectation value of the
quantum analysis, but both the description and the dispersion are completely
different.\cite{AJP} \ 

In both the classical and the quantum theories, the $SO\left(  2\right)  $
symmetry involving $x$ and $p$ is recognized by some authors. \ However, in
elementary quantum texts, the symmetry is often not
acknowledged.\cite{Griffiths} \ The $SO\left(  2\right)  $ symmetry involving
$x$ and $p$ alone leads to the integer-indexed representations\cite{Miller}
and to the average values. \ The fluctuations, however, are very different
between the classical and quantum theories. \ 

In any case, it seems comforting to those with classical sensibilities that
some parts of old quantum theory can be understood as classical charges moving
in classical trajectories under the fluctuations of random classical
zero-point radiation.

\section{Concluding Remarks}

\subsection{Resonance: Its Absence in Classical Statistical Mechanics and Its
Importance in Classical Electromagnetism}

At the end of the 19th century and beginning of the 20th, there were repeated
attempts to apply nonrelativistic classical statistical mechanics to phenomena
associated with atomic physics. \ The Rayleigh-Jeans law is the result of such
an attempt. \ Indeed, Planck investigated blackbody radiation from the
perspective of thermodynamics, and connected the average energy of a charged
harmonic oscillator with the average energy of the classical radiation modes
at the same frequency as the harmonic oscillator.\ \ However, he did not
introduce classical electromagnetic zero-point radiation. \ 

There is a huge difference between the Brownian motion treated in texts of
statistical mechanics and the motion of a charged mechanical system in
classical electrodynamics with classical electromagnetic zero-point radiation.
\ For example, due to collisions, a \textit{neutral} (uncharged)
one-dimensional mechanical harmonic oscillator with natural frequency
$\omega_{0}$ will come to \textit{thermal} equilibrium at $k_{B}T,$
\textit{independent} of the oscillator frequency $\omega_{0}$. \ On the other
hand, in electromagnetism, the response of a charged oscillator is resonant at
its natural frequency $\omega_{0}$. \ Indeed, the forcing of a charged
oscillator by an electric field involves the location of the charged particle,
$\mathbf{F}\left(  t\right)  =e\mathbf{E}\left[  \mathbf{r}_{e}\left(
t\right)  ,t\right]  $, so that the amplitude of the oscillation
$\mathbf{r}_{e}\left(  t\right)  $ can influence the driving electromagnetic
force on the oscillator. In the \textit{ground state}, the higher multiples of
the classical zero-point \textit{radiation} field lead to the same stochastic
behavior for the charged oscillator as given by the dipole approximation.
\ There is no change in the spectrum of classical zero-point radiation due to
the charged harmonic oscillator in the small-source approximation. \ However,
the dependence of the electromagnetic force on the amplitude of the
oscillation will lead to a parametric forcing which leads to resonant excited
states for a charged one-dimensional linear oscillator. \ In this article, we
have used classical electrodynamics with classical electromagnetic zero-point
radiation to describe the motion of a \textit{charged} harmonic oscillator in
zero-point radiation. \ \ 

\subsection{The Ground State in Zero-Point Radiation}

For the one-dimensional charged harmonic oscillator \textit{ground} state
involving $l=1$, \textit{dipole} radiation \textit{determines} the amplitude
$Z_{0}$ of mechanical oscillation in terms of \textit{any} spectrum of random
radiation. \ The charged harmonic oscillator will come to equilibrium in
\textit{any} arbitrary spectrum of random radiation. \ On the other hand, if
we require that classical electrodynamics holds and the radiation spectrum is
in \textit{equilibrium} for \textit{both} dipole \textit{and} quadrupole
radiation, and any velocity-dependent damping for the oscillator is omitted,
then (up to an overall multiplicative constant) the \textit{only} allowed
spectrum of random radiation is that of Lorentz-invariant classical zero-point
radiation. \ \textit{In classical electromagnetic zero-point radiation,}
\textit{both the dipole and quadrupole radiation balance involve the same
stochastic process for the charged oscillator. \ }However, the quadrupole
radiation is suppressed by additional powers of the speed of light $1/c^{2}$.
Based upon the ground state radiation behavior, one would be unaware of the
presence of zero-point radiation for a charged harmonic oscillator in its
ground state. \ 

\subsection{Resonant Excited States and Dipole Radiation}

For any resonant \textit{excited} states, we do \textit{not} expect all the
radiation modes to contribute to the same stochastic process for the charged
oscillator since excited states are unstable and decay. \ Therefore we focus
out attention on the \textit{dipole} radiation associated with possible
excited states. \ In the nonrelativistic calculation for the charged
mechanical system, dipole radiation is the predominant radiation multipole.
\ Since zero-point radiation is Lorentz-invariant while the harmonic
oscillator potential is not, compatibility requires that the velocity (and
hence any amplitude of the oscillation) is very small. \ Now there is only one
steady-state resonant frequency for the mechanical harmonic oscillator, namely
its natural oscillation frequency $\omega_{0}$. \ However, because of the
\textit{position dependence }$\mathbf{E}\left[  \widehat{z}z\left(  t\right)
,t\right]  $\textit{ of the driving radiation}, and hence of the driving
force, the resonant excited states involve basically the $l=1,m=0,$ spherical
multipole fields, but for \textit{different frequencies between the emitted
radiation and the driving radiation. \ }For the one-dimensional charged
harmonic oscillator, the radiation \textit{emission} is always at the natural
oscillation frequency $\omega_{0}$, but, for resonant excited states, the
oscillation amplitude $Z_{n}$ is larger, whereas the \textit{driving
radiation} will be at $\omega_{rad-n}$. \ What we require is that the
\textit{net driving force contained in }$E\left[  \widehat{z}z\left(
t\right)  ,t\right]  $\textit{ contains a frequency component agreeing with
the natural frequency of the mechanical oscillator }$\omega_{0}$. \ We find
that the oscillator \textit{energy }$\mathcal{E}_{e-n}=J_{e-n}\omega
_{0}=\left[  \left(  2n+1\right)  J_{e-0}\right]  \omega_{0}$ of the charged
oscillator\ (oscillating at $\omega_{0}$) is the same as the \textit{energy
}$\mathcal{E}_{rad-n}=J_{rad}\omega_{rad-n}=J_{rad}\left[  \left(
2n+1\right)  \omega_{0}\right]  $\textit{ } of the driving zero-point
radiation mode (of frequency $\omega_{rad-n}=(2n+1)\omega_{0}$). \ On change
of the excited state, the energy change of the oscillator is the same as the
energy change of the radiation modes, but the emitted radiation corresponds to
that of a charged oscillator in empty space. \ The important role of classical
zero-point radiation in setting the exact average amplitudes for both the
ground state and the resonant excited states seems hidden. \ The present work
provides a classical electromagnetic understanding for the old-quantum picture
of electrons in classical orbits. \ 

\section{Acknowledgements}

I am deeply indebted to the work of Professor Daniel C. Cole whose article
with Y. Zou kept the analysis of classical zero-point radiation advancing when
only linear systems seemed successful, and whose work on subharmonic
resonances was quite thought-provoking. \ The work by Professor Herman
Batelaan and W. Huang was also intriguing. \ Furthermore, there are many books
and articles on classical electrodynamics, quantum mechanics, quantum field
theory, group theory, and statistical physics which have been instrumental in
my understanding of the classical electromagnetic situation. \ I also wish to
thank Professor Cole for noting typos and ambiguous aspects of the manuscript. \ \

\bigskip

March 1, 2025 \ \ \ \ \ \ 1D-SHOz15.tex

\end{document}